\documentclass[10pt, conference]{IEEEtran}
\usepackage{cite}
\usepackage{amsmath,amssymb,amsfonts}
\usepackage{graphicx}
\usepackage[colorlinks]{hyperref}
\usepackage{xurl}
\usepackage{cleveref}
\usepackage{xcolor} %
\def\tmp#1#2#3{%
  \definecolor{Hy#1color}{#2}{#3}%
  \hypersetup{#1color=Hy#1color}}
\tmp{link}{HTML}{800006}
\tmp{cite}{HTML}{2E7E2A}
\tmp{file}{HTML}{131877}
\tmp{url} {HTML}{8A0087}
\tmp{menu}{HTML}{727500}
\tmp{run} {HTML}{137776}
\def\tmp#1#2{%
  \colorlet{Hy#1bordercolor}{Hy#1color#2}%
  \hypersetup{#1bordercolor=Hy#1bordercolor}}
\tmp{link}{!60!white}
\tmp{cite}{!60!white}
\tmp{file}{!60!white}
\tmp{url} {!60!white}
\tmp{menu}{!60!white}
\tmp{run} {!60!white}

\usepackage{booktabs}
\usepackage[inline]{enumitem}
\usepackage{xcolor}
\usepackage{multirow}
\usepackage{makecell}

\usepackage{pdflscape}

\begin{document}

\title{Improving Zero-Noise Extrapolation \\ via Physically Bounded Models}

\author{
\IEEEauthorblockN{Andriy Miranskyy, Adam Sorrenti, Jasmine Thind, Claude Gravel}
\IEEEauthorblockA{\textit{Department of Computer Science, Toronto Metropolitan University} \\
Toronto, Canada \\
\{avm, adam.sorrenti, j1thind, gravel\}@torontomu.ca
}}

\maketitle

\begin{abstract}
Zero-noise extrapolation (ZNE) mitigates errors in near-term quantum devices by extrapolating measurements obtained at amplified noise levels to estimate noise-free expectation values. In practice, commonly used extrapolation models are fitted without enforcing physical constraints, which can yield predictions outside the valid range of quantum observables.

In this work, we introduce physically bounded variants of polynomial, exponential, and polynomial--exponential extrapolation models by explicitly parameterizing the zero-noise estimate and constraining it during optimization. We evaluate the approach using a large synthetic benchmark comprising 180{,}000 circuits and approximately 3.6 million ZNE experiments generated under realistic device noise models derived from IBM quantum backends. We also perform preliminary validation on real quantum hardware using GHZ and W-state circuits.

Across the synthetic benchmark, bounded extrapolation substantially reduces unphysical predictions and improves the stability of exponential- and polynomial--exponential-family models, whereas polynomial models show little difference between bounded and unbounded variants. Hardware experiments show similar qualitative behaviour: bounded models generally avoid pathological extrapolations and often provide a more reliable balance between accuracy and usable coverage. At the same time, the results highlight practical limitations of current devices, including stronger-than-expected noise effects and variability not fully captured by simulation models. These results suggest that enforcing physical constraints during extrapolation improves the reliability of ZNE and that this approach can be incorporated into existing workflows with minimal modification.

\end{abstract}

\section{Introduction}

Quantum computing devices available today operate in the noisy intermediate-scale quantum (NISQ) regime, where decoherence, gate imperfections, and measurement errors significantly limit computational accuracy~\cite{preskill2018quantum}. Although quantum hardware has improved steadily in recent years, error rates remain high enough that many quantum algorithms cannot yet be executed reliably without additional noise-mitigation techniques~\cite{bharti2022noisy}. As a result, quantum error mitigation has emerged as a critical research direction to enable useful computation on near-term devices (see~\cite{cai2023quantum} for a review).

Zero-noise extrapolation (ZNE) is one of the most widely used error-mitigation techniques for NISQ systems~\cite{temme2017error,li2017efficient}. The central idea of ZNE is to artificially amplify circuit noise and measure the observable of interest at several noise levels. These noisy measurements are then extrapolated to estimate the value that would be obtained in the absence of noise. ZNE has the advantage of requiring no additional qubits or full fault-tolerant error correction, and it can be applied to a wide range of quantum circuits and observables.

In practice, ZNE relies on fitting a parametric model to measurements taken at different noise scale factors. Commonly used models include polynomial, exponential, and polynomial--exponential extrapolators~\cite{giurgica2020digital}. While these approaches have proven effective in many settings, they suffer from an important limitation: the fitted models are typically unconstrained and may therefore yield extrapolated predictions that violate the physical limits of quantum observables~\cite{larose2022mitiq}. For expectation values of $\pm 1$-valued observables, the physically admissible range is $[-1,1]$. Nevertheless, unconstrained extrapolators can produce predictions outside this interval, particularly when the number of available noise levels is small or when measurements are strongly affected by noise. Such unphysical predictions may degrade mitigation accuracy and introduce instability into the extrapolation process.

In this work, we investigate whether enforcing physically valid bounds during extrapolation can improve the reliability and accuracy of ZNE. We introduce \emph{physically bounded extrapolation models} that explicitly constrain the zero-noise prediction to the valid range of expectation values. Our approach reformulates several commonly used extrapolation models so that the zero-noise value appears directly as a model parameter, enabling simple bound constraints to be imposed during optimization. This ensures that the extrapolated value remains physically valid while preserving the flexibility of the underlying parametric models.

To evaluate the proposed approach, we construct a large-scale synthetic benchmark of quantum circuits spanning a broad range of expectation values and circuit structures. The dataset comprises $180{,}000$ circuits and $\approx 3.6$ million ZNE experiments generated under realistic device noise models derived from a modern suite of IBM quantum backends. Using this benchmark, we compare bounded and unbounded variants of several extrapolation models and analyse their behaviour across expectation-value regimes. Our results show that bounded extrapolation substantially reduces unphysical predictions and consistently improves mitigation accuracy.

We further observe that the benefits of physical bounding vary by extrapolation family: polynomial models are often only weakly affected, whereas exponential-family models show substantially improved robustness and accuracy. Finally, we present preliminary validation of the proposed methods on real quantum hardware using analytically tractable circuits from the MQTBench benchmark suite~\cite{quetschlich2023mqt}.

The \textit{contributions} of this work are as follows:
\begin{itemize}
\item We introduce physically bounded variants of common ZNE extrapolation models, including exponential, polynomial, and polynomial--exponential formulations.
\item We construct a large-scale synthetic benchmark of quantum circuits using realistic device noise models and use it to systematically evaluate bounded and unbounded extrapolation models.
\item We show empirically that bounded extrapolation reduces unphysical predictions and improves the robustness and mitigation accuracy of exponential- and polynomial--exponential-family models compared with conventional unconstrained approaches.
\item We show that polynomial models are usually only weakly affected by bounds, helping clarify when physical constraints matter most in practice.
\item We provide preliminary validation of the approach on real quantum hardware using circuits from MQTBench.
\end{itemize}

To facilitate reproducibility and further research on error mitigation, we \textit{release} the full dataset of circuits and measurement results~\cite{miranskyy2026dataset}, along with the reference implementations of the extrapolation models used in this study at~\cite{miranskyy2026code}.

The remainder of this paper is organized as follows. \Cref{sec:background} reviews the principles of zero-noise extrapolation and commonly used extrapolation models. \Cref{sec:bounded} introduces the proposed physically bounded extrapolation formulations. \Cref{sec:synthetic} describes the synthetic benchmark dataset and presents the main experimental results. %
\Cref{sec:hardware} presents preliminary validation on real quantum hardware. Finally, \Cref{sec:discussion} discusses limitations and future directions, and \Cref{sec:conclusion} concludes the paper.

\section{Background}
\label{sec:background}

\subsection{Zero-Noise Extrapolation}

Zero-noise extrapolation (ZNE) is a widely used technique for mitigating errors in near-term quantum devices. The central idea is to estimate the expectation value of an observable in the absence of noise by measuring the same circuit under artificially amplified noise and extrapolating the results to the zero-noise limit.

Let $E(0)$ denote the ideal expectation value of an observable, and $E(\lambda)$ the expectation value measured at a noise scale factor $\lambda \ge 1$ ($\lambda = 1$ corresponds to the original circuit). The parameter $\lambda$ is a \emph{noise amplification factor} that scales the effective noise experienced by the circuit while preserving its logical functionality. Given a quantum circuit $C$, ZNE evaluates a sequence of modified circuits $C(\lambda_1), C(\lambda_2), \dots, C(\lambda_n)$ with increasing noise levels $\lambda_1, \lambda_2, \dots, \lambda_n$. The resulting measurements yield a set of expectation values $E(\lambda_1), E(\lambda_2), \dots, E(\lambda_n)$. A parametric model $\hat{E}(\lambda)$ is then fitted to these measurements, and the extrapolated value $\hat{E}(0)$ is used as an estimate of the zero-noise expectation value.

In practice, noise amplification is typically implemented using \emph{gate folding}. If $U$ is a unitary gate, the transformation $U \rightarrow U U^\dagger U$ preserves the ideal computation while increasing the effective noise experienced by the circuit. Applying this transformation throughout the circuit yields executions with larger noise scale factors $\lambda$. For example, $\lambda = 3$ can be obtained by replacing each gate $U$ with $U U^\dagger U$. Fractional scale factors, such as $\lambda = 1.5$ or $\lambda = 2$, can be achieved by applying the folding transformation to only a subset of gates in the circuit; see~\cite{giurgica2020digital} for details.

Assuming the expectation value varies smoothly with the effective noise strength, the function $E(\lambda)$ can be approximated by a parametric model. Extrapolating this model to $\lambda = 0$ yields an estimate of the ideal, noise-free expectation value.

\subsection{Extrapolation Models}\label{sec:orig_models}

The accuracy of ZNE depends strongly on the model used to extrapolate noisy measurements to the zero-noise limit. A variety of extrapolation models have been proposed in the literature, ranging from simple analytic functions motivated by noise physics to more flexible regression-based approaches. The formulas below follow the definitions used in the Mitiq v.0.48.1 package~\cite{larose2022mitiq} (definitions vary across the literature).

\paragraph{Polynomial extrapolation}

A common approach~\cite{giurgica2020digital} is to model the expectation value as a polynomial function of the noise scale factor:
\begin{equation}\label{eq:poly_model}
\hat{E}(\lambda) = \theta_0 + \theta_1 \lambda + \theta_2 \lambda^2 + \cdots + \theta_d \lambda^d .
\end{equation}

The zero-noise estimate is then given by $\hat{E}(0) = \theta_0$. Polynomial extrapolation is simple to implement but may be sensitive to noise and overfitting when higher-degree polynomials are used.

\paragraph{Exponential extrapolation}

Another widely used model~\cite{endo2018practical} assumes exponential decay of the expectation value with increasing noise:
\begin{equation}\label{eq:exp_model}
\hat{E}(\lambda) = a + b e^{-c \lambda},
\end{equation}
where $a$, $b$, and $c$ are model parameters. This form is motivated by simple noise processes in which expectation values decay approximately exponentially with circuit depth or effective noise strength.

\paragraph{Polynomial--exponential extrapolation}

A commonly used extrapolation model~\cite{giurgica2020digital} assumes that the expectation value follows an exponential trend whose exponent is itself a polynomial function of the noise scale factor. In this formulation, the model is written as
\begin{equation}\label{eq:polyexp_model}
\hat{E}(\lambda) =
a + \mathrm{sign} \cdot \exp\left(z(\lambda)\right),
\end{equation}
where $a$ is an asymptotic value, $\mathrm{sign} \in \{ -1, 1 \}$ determines whether the exponential is increasing or decreasing, and $z(\lambda)$ is a polynomial of order $d$,

\begin{equation}\label{eq:polyexp_exponent}
z(\lambda) =
c_0 + c_1 \lambda + c_2 \lambda^2 + \cdots + c_d \lambda^d .
\end{equation}

The sign parameter allows the model to capture either increasing or decreasing exponential behaviour and is typically inferred from the observed data. This formulation provides a flexible functional form capable of modelling a wide range of noise-dependent behaviours.

\subsection{Physical Constraints on Expectation Values}

For many quantum observables, particularly Pauli operators, measurement outcomes take values in $\{-1,1\}$. Consequently, the expectation value of such observables must lie within the interval $-1 \le E \le 1$. In practice, however, standard ZNE extrapolators are typically fit (e.g., in popular Mitiq~\cite{larose2022mitiq} and ZNE Prototype~\cite{zneprototype,majumdar2023best} packages) without enforcing this physical constraint. When measurements are noisy or only a small number of noise levels are available, the extrapolated prediction may fall outside the valid range of expectation values. This behaviour is particularly common with higher-order polynomial fits or poorly conditioned exponential models. Predictions outside the physically valid range can yield inaccurate mitigation results and may signal instability in the extrapolation procedure. This observation motivates the development of extrapolation models that explicitly enforce the physical constraints of quantum observables during parameter estimation. In the following section, we introduce bounded variants of the models introduced in \Cref{sec:orig_models} that enforce the constraint $-1 \le E(0) \le 1$ during optimization.

\section{Physically Bounded Extrapolation}
\label{sec:bounded}

The extrapolation models introduced in \Cref{sec:orig_models}, including the polynomial model in \Cref{eq:poly_model}, the exponential model in \Cref{eq:exp_model}, and the polynomial--exponential formulation in \Cref{eq:polyexp_model}, are typically fitted without enforcing physical constraints on the extrapolated value. For many observables of interest in quantum algorithms, particularly Pauli operators, measurement outcomes lie in $\{-1,1\}$, and the corresponding expectation values must therefore satisfy
\begin{equation*}
-1 \le E \le 1 .
\end{equation*}

However, unconstrained extrapolation models may yield predictions outside this interval, especially when only a small number of noisy observations are available or the fitted curve is poorly conditioned. Such unphysical predictions can reduce the accuracy of error mitigation and introduce instability into the extrapolation process. To address this issue, we introduce \emph{physically bounded extrapolation models} that explicitly enforce the constraint
\begin{equation*}
-1 \le \hat{E}(0) \le 1
\end{equation*}
during parameter estimation. The key idea is to parameterize each extrapolation model so that the zero-noise prediction appears explicitly as a model parameter, which can then be constrained during optimization.

Let $\{(\lambda_i, y_i)\}_{i=1}^{N}$ denote the measured expectation values $y_i = E(\lambda_i)$ obtained at noise scale factors $\lambda_i$, where $N$ is the number of noise scale factors (i.e., the number of measurement points used for extrapolation).
Each bounded model is fitted by solving a constrained least-squares problem
\begin{equation*}
\min_{\Theta}
\sum_{i=1}^{N}
\left[y_i - \hat{E}(\lambda_i; \Theta)\right]^2 ,
\end{equation*}
subject to constraints ensuring that the extrapolated zero-noise value lies within the physically valid interval. Here, $\Theta$ denotes the vector of model parameters that define the functional form of the extrapolation model $\hat{E}(\lambda; \Theta)$.  In our implementation, the resulting optimization problems are solved using the L-BFGS-B algorithm~\cite{byrd1995limited}  (using the default parameters from the implementation in SciPy v.1.16.2).\footnote{We also evaluated a nonlinear least-squares implementation based on SciPy v.1.16.2 \texttt{curve\_fit} function~\cite{virtanen2020scipy}, but found its performance to be inferior to that of the constrained L-BFGS-B-based optimization used in this work.}

\subsection{Bounded Polynomial Extrapolation}

We first consider a bounded variant of the polynomial extrapolation model in \Cref{eq:poly_model}. The polynomial model is
\begin{equation}
\hat{E}(\lambda) =
\theta_0 + \theta_1\lambda + \theta_2\lambda^2 + \cdots + \theta_d\lambda^d .
\end{equation}

Since the zero-noise prediction corresponds to the intercept,
$
\hat{E}(0) = \theta_0 ,
$
we impose the physical constraint directly on this parameter. The resulting optimization problem is
\begin{equation*}
\min_{\theta_0,\ldots,\theta_d}
\sum_{i=1}^{N}
\left[
y_i -
\sum_{j=0}^{d}\theta_j \lambda_i^j
\right]^2
\end{equation*}
subject to
$
-1 \le \theta_0 \le 1 .
$
The remaining polynomial coefficients are left unconstrained.

\subsection{Bounded Exponential Extrapolation}

We next consider a bounded variant of the exponential model introduced in \Cref{eq:exp_model}. To make the zero-noise value explicit, we reparameterize the model as follows
\begin{equation}\label{eq:bounded_exp}
\hat{E}(\lambda) =
a + (\zeta - a)e^{-c\lambda},
\end{equation}
where $a$ represents the asymptotic value at large noise levels, $c>0$ controls the decay rate, and $\zeta$ denotes the extrapolated zero-noise value. This parameterization ensures that
$
\hat{E}(0) = \zeta .
$
The parameters are obtained by solving
\begin{equation*}
\min_{a,\zeta,c}
\sum_{i=1}^{N}
\left[
y_i -
\left(
a + (\zeta - a)e^{-c\lambda_i}
\right)
\right]^2
\end{equation*}
subject to
$
-1 \le a \le 1, \, -1 \le \zeta \le 1, \, c > 0 .
$
When prior knowledge about the infinite-noise asymptote is available, $a$ can optionally be fixed to a predefined value (for example, $a = 0$ for traceless observables), reducing the number of free parameters~\cite{larose2022mitiq}.

\subsection{Bounded Polynomial--Exponential Extrapolation}

Finally, we construct a bounded variant of the polynomial--exponential model introduced in \Cref{eq:polyexp_model}. We reparameterize the model as
\begin{equation}\label{eq:bounded_poly_exp}
\hat{E}(\lambda) =
a + (\zeta - a)e^{r(\lambda)},
\end{equation}
where the exponent is modelled as a polynomial
\begin{equation}\label{eq:r_lambda}
r(\lambda) =
c_1\lambda + c_2\lambda^2 + \cdots + c_d\lambda^d .
\end{equation}
This formulation\footnote{
In \Cref{eq:r_lambda}, the constant term $c_0 = 0$ is intentionally omitted so that $r(0)=0$, ensuring that the model satisfies $\hat{E}(0)=\zeta$.} again makes the zero-noise value explicit, $\hat{E}(0) = \zeta$. The parameters are obtained by solving
\begin{equation*}
\min_{a,\zeta,c_1,\ldots,c_d}
\sum_{i=1}^{N}
\left[
y_i -
\left(
a + (\zeta - a)
\exp\left(
\sum_{j=1}^{d} c_j \lambda_i^j
\right)
\right)
\right]^2
\end{equation*}
subject to
$-1 \le a \le 1, \, -1 \le \zeta \le 1 .$ Note that due to differing constraints, \Cref{eq:bounded_exp,eq:bounded_poly_exp} are not necessarily equal when $d = 1$.

\subsection{Discussion}

The bounded formulations preserve the functional forms of the extrapolation models described in \Cref{sec:background} while explicitly constraining the zero-noise value during optimization. Unlike post hoc clipping of extrapolated predictions, the bound is enforced directly during model fitting, allowing the physical constraint to influence the optimization process. As we demonstrate in \Cref{sec:synthetic}, this modification eliminates unphysical predictions and improves mitigation accuracy across a wide range of circuits and noise regimes.

\section{Synthetic Benchmark Evaluation}
\label{sec:synthetic}

To evaluate the effectiveness of the proposed bounded extrapolation models, we construct a large-scale synthetic benchmark that covers a broad range of circuits, expectation values, and noise regimes. The goal of this dataset is to provide a controlled yet realistic environment for analyzing the behaviour of ZNE extrapolation models across a range of experimental configurations.

\subsection{Synthetic Circuit Generation}

We generate a large collection of quantum circuits with qubit counts $n$ ranging from 2 to 10 and target circuit depths in the set $\{10, 20, 30, 40, 50\}$. Circuits are generated randomly using the gate set available in Qiskit v.2.3.0~\cite{javadi2024quantum} and subsequently transpiled without optimization into the basis \textsc{\{u1, u2, u3, cx\}}. This approach avoids depth-reduction optimizations that could otherwise obscure the relationship between circuit structure and accumulated noise.

To ensure broad coverage of expectation values, we employ a bucketization strategy based on the ideal expectation value of the observable. Specifically, the interval $[-1,1]$ is partitioned into bins of width $0.05$, and circuits are generated using rejection sampling until 100 circuits per bin are obtained.

\begin{figure}[tb]
\centering
\includegraphics[width=0.9\linewidth]{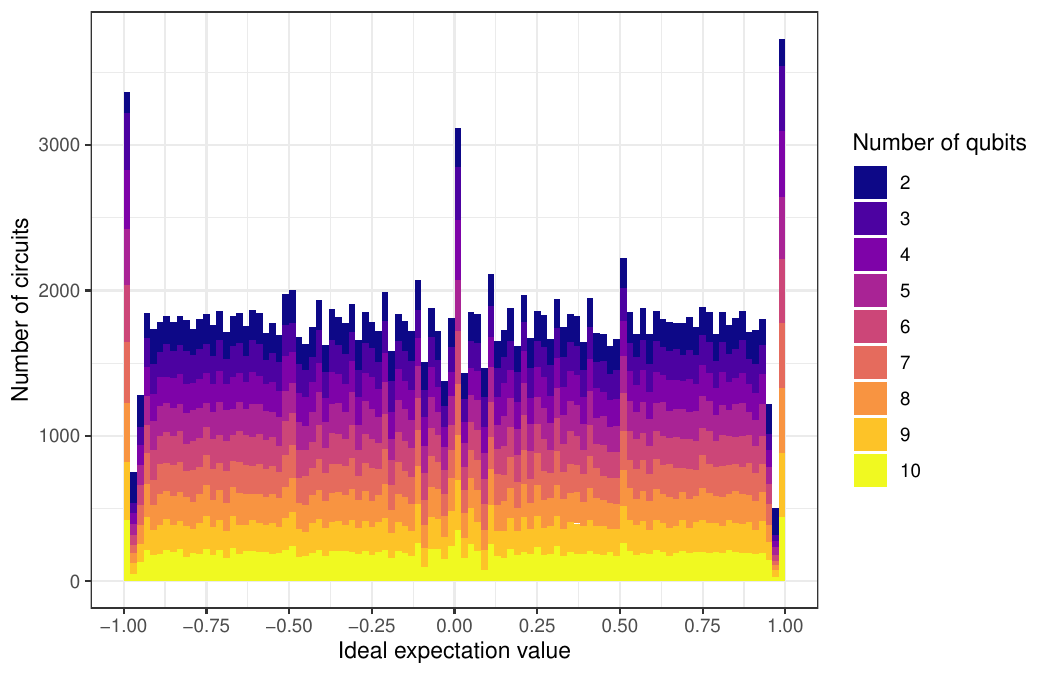}
\caption{Histogram of ideal expectation values in the synthetic benchmark (bin width $0.025$, chosen to highlight clustering near values, such as $-1$, $0$, $1$). }
\label{fig:expectation_distribution}
\end{figure}

The distribution of ideal expectation values across all generated circuits is shown in \Cref{fig:expectation_distribution}.
Overall, the resulting dataset covers the interval $[-1,1]$ approximately uniformly. However, random circuit generation produces a higher concentration of cases near special expectation values, such as $-1$, $0$, and $1$, where observables become strongly aligned, anti-aligned, or cancel due to symmetry.

The ideal expectation values are computed using state vector simulation. For a circuit $C$ and observable $O$, the ideal value is evaluated as $E^{\mathrm{ideal}} = \langle \psi | O | \psi \rangle,$ where $|\psi\rangle$ denotes the state produced by executing the circuit on a noiseless simulator. Using this procedure, we construct a dataset containing $180{,}000$ unique circuits (computed as $40$ observable-value bins $\times$ $100$ circuits per bin $\times$ $5$ circuit depths $\times$ $9$ qubit counts) spanning a wide range of circuit structures.

\subsection{Noise Model and Simulation}

To simulate hardware noise, we use device noise models derived from IBM quantum backends. Specifically, we employ the fake device models \texttt{FakeAlgiers} and \texttt{FakeFez}, corresponding to IBM devices from the 27-qubit Falcon r5. 11 and 156-qubit Heron~r2 processor families, respectively~\cite{javadi2024quantum}. Circuits are transpiled into the basis gates native to each device before simulation, thereby capturing the distinct noise characteristics and connectivity constraints of these architectures.

For computational efficiency, we simulate only the subset of qubits required by each circuit rather than the full device topology. This simplification makes large-scale classical simulation feasible, but yields noise behaviour that is more optimistic than that observed when executing the same circuits on the full physical devices. In particular, interactions with unused qubits and other global hardware effects are not captured. Nevertheless, this approach enables extensive controlled experimentation across millions of ZNE instances while preserving the key device-specific noise parameters.

\subsection{Zero-Noise Extrapolation Configuration}

For each circuit in the dataset, we perform ZNE experiments using gate folding to amplify noise. We evaluate three sets of noise scale factors: $\lambda \in \{1,2,3\}, \,  \{1,3,5\}, \, \{1,2,3,4,5\}$.

Each circuit is executed with $10{,}000$ measurement shots, and the experiment is repeated ten times to capture statistical variability. We treat these repetitions as independent measurement instances. The resulting measurement outcomes are used to compute the expectation value of the observable at each noise scale factor.

The nominal total is $3.6$ million experiments (computed as $180{,}000$ circuits $\times$ $2$ device noise models $\times$ $10$ repetitions); however, a small number ($431$, or $0.01$\%) of measurements are excluded due to rounding errors. As a result, the final dataset contains $3{,}599{,}569$ ZNE experiment instances across $179{,}979$ unique circuits.

\subsection{Evaluated Models}
Table~\ref{tab:evaluated_models} summarizes the three extrapolation families that form the focus of this study: polynomial, exponential, and polynomial--exponential models. For each family, we evaluate both conventional unbounded formulations and physically bounded counterparts. Conceptual descriptions of the model families\footnote{We also evaluated several additional baselines, including Richardson extrapolation (comparable to polynomial in Mitiq implementation)~\cite{larose2022mitiq}, fake-nodes extrapolation~\cite{de2020polynomial}, variance-aware methods~\cite{zneprototype}, and regressors such as ridge regression~\cite{hoerl1970ridge}, LASSO regression~\cite{tibshirani1996regression}, support vector regression~\cite{cortes1995support}, decision trees~\cite{breiman1984classification}, random forests~\cite{breiman2001random}, and XGBoost~\cite{chen2016xgboost}. These alternatives exhibited performance comparable to or weaker than that of the polynomial, exponential, and polynomial--exponential families in our synthetic benchmark and are omitted from the main presentation for brevity.}
are given in \Cref{sec:orig_models}. The unbounded baselines were used as implemented in Mitiq v0.48.1~\cite{larose2022mitiq}, while the bounded variants were implemented via constrained optimization using SciPy v1.16.2~\cite{virtanen2020scipy}.

When only three noise amplification factors $\lambda$ are available, some higher-capacity models cannot be fitted because the number of parameters exceeds the number of observations. For example, a degree $d=3$ polynomial requires four parameters and therefore cannot be estimated from three data points. The same limitation applies to the polynomial--exponential model with asymptote $a=0$ and $d=3$, as well as to the polynomial--exponential model with a free asymptote and $d=2$ or $3$. Therefore, these model variants are excluded from experiments using the three-$\lambda$ configuration.

\begin{table}[tb]
\centering
\caption{Core extrapolation model families evaluated in this work.}
\label{tab:evaluated_models}
\setlength{\tabcolsep}{4pt}
\renewcommand{\arraystretch}{1.1}
\begin{tabular}{@{}p{0.18\linewidth}p{0.75\linewidth}@{}}
\toprule
\textbf{Family} & \textbf{Variants considered} \\
\midrule
Polynomial
& Bounded and unbounded; degree $d \in \{1,2,3\}$ \\
Exponential
& Bounded and unbounded; asymptote $a \in \{\text{estimated}, 0\}$ \\
Polynomial--exponential
& Bounded and unbounded; degree $d \in \{1,2,3\}$; asymptote $a \in \{\text{estimated}, 0\}$ \\
\bottomrule
\end{tabular}
\end{table}

\subsection{Dataset Overview}

\begin{figure}[tb]
\centering
\includegraphics[width=\linewidth]{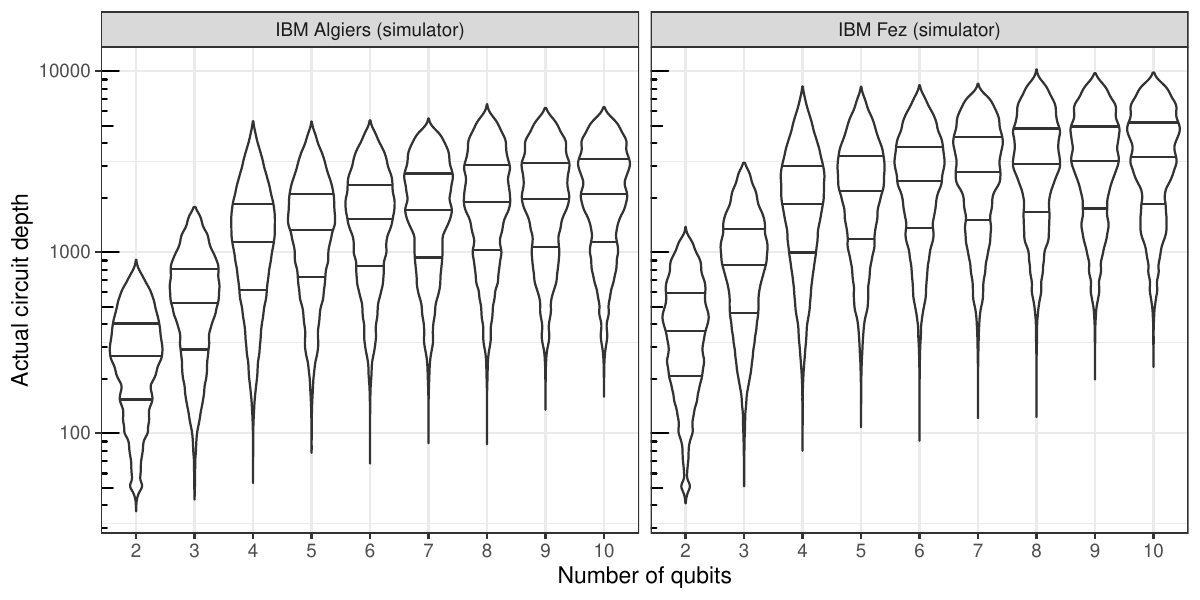}
\caption{Distribution of transpiled circuit depths in the synthetic benchmark. Although the target circuit depths range from 10 to 50 gates, device-aware transpilation can substantially increase them. Vertical bars indicate the 25th, 50th (median), and 75th quartiles.}
\label{fig:depth_distribution}
\end{figure}

The resulting benchmark dataset spans a wide range of circuit sizes, noise regimes, and expectation values. Because circuits are initially generated with target depths between 10 and 50 and then transpiled to device-specific gate sets that accommodate the hardware qubit topology, the final circuit depths vary substantially. In some cases, they can exceed thousands of gates, as shown in \Cref{fig:depth_distribution}. This diversity of circuit structures allows us to evaluate extrapolation models across a wide range of noise conditions.

\subsection{Aggregate Model Comparison}

We evaluate the accuracy of each extrapolation model using two primary metrics: mean absolute error (MAE) and mean squared error (MSE) relative to the ideal expectation values obtained from statevector simulations. Formally, given $M$ circuit instances with ideal expectation values $E^{\mathrm{ideal}}_j$ and mitigated estimates $\hat{E}_j$, the metrics are defined as
\begin{equation}\label{eq:mae}
\mathrm{MAE} = \frac{1}{M}\sum_{j=1}^{M} {\left| \hat{E}_j - E^{\mathrm{ideal}}_j \right|},
\end{equation}
and
\begin{equation}\label{eq:mse}
\mathrm{MSE} = \frac{1}{M}\sum_{j=1}^{M}{\left( \hat{E}_j - E^{\mathrm{ideal}}_j \right)^2}.
\end{equation}
Here, each instance $j$ corresponds to a single repetition of a retained circuit--noise-model--scale-factor configuration, treated as an independent measurement.

\paragraph{Convergence and mean error}
The aggregate performance of the evaluated models is summarized in Appendix~\ref{app:model_comparison}  (\Cref{tab:paired_bounded_unbounded_summary}). Two principal patterns emerge.

First, bounded and unbounded polynomial models exhibit nearly identical behaviour. Polynomial fits rarely extrapolate outside the physically valid interval $[-1,1]$ in this benchmark, so constraining the zero-noise estimate has no material effect.

Second, exponential-family models show clear differences. When the asymptote is treated as a free parameter, optimization frequently becomes unstable for unbounded models: they converge in only approximately 70\%--86\% of cases, while bounded formulations converge in more than 99.6\% of instances. Fixing the asymptote to a physically motivated value (e.g., $a=0$ for traceless observables) substantially stabilizes both formulations, with bounded models achieving convergence rates above 99.8\% and unbounded models converging in nearly all cases. Across all configurations in which both variants converge, bounded models consistently achieve lower MAE and MSE.

Among all evaluated models, the best overall performance is achieved by low-capacity formulations that incorporate physical assumptions about the observable's asymptotic behaviour. In particular, the degree $d=1$ polynomial--exponential model with asymptote $a=0$ offers the strongest balance of robustness and accuracy in the synthetic benchmark.

\paragraph{Statistical significance and effect sizes}
To verify that the observed differences are not artifacts of aggregate statistics, we perform paired comparisons on matched experimental instances. Each bounded model is compared only with its corresponding unbounded analogue under identical experimental conditions, with comparisons grouped by noise scale factors, quantum processor, and number of qubits.

For each matched pair, we compute
\begin{equation} \label{eq:improvement}
\Delta = |\hat E^{\mathrm{unbounded}} - E^{\mathrm{ideal}}| - |\hat E^{\mathrm{bounded}} - E^{\mathrm{ideal}}|,
\end{equation}
so that $\Delta > 0$ indicates improved accuracy of the bounded model. Statistical significance of paired differences is assessed using the two-sided Wilcoxon signed-rank test~\cite{wilcoxon1945individual}, with Holm correction for multiple comparisons~\cite{holm1979simple} at $\alpha=0.05$. Effect sizes are reported as Cohen's $d$ for paired differences~\cite{cohen1988statistical}.

For polynomial models, differences between bounded and unbounded formulations are negligible, consistent with the observation above that they rarely violate the physical bounds. For exponential and polynomial--exponential families, the paired Wilcoxon test detected a statistically significant difference in 99.5\% of matched-group comparisons (Holm-adjusted $p<0.05$); of these, 94\% favoured the bounded model and 6\% the unbounded one. For each matched-group comparison, Cohen's $d$ was computed from the paired absolute-error differences defined in \Cref{eq:improvement}. Among comparisons that remained significant after Holm correction, 90\% had negligible effect sizes and the remaining 10\% had small effect sizes. Given that both methods are already reasonably accurate on much of the synthetic benchmark, a small but consistent effect favouring the bounded model across millions of matched experiments is a meaningful practical signal; moreover, as discussed below, these paired statistics exclude configurations in which the unbounded models fail outright.

Interpret the paired statistics conservatively. With an unconstrained asymptote, unbounded exponential-family models often fail to converge or produce non-finite predictions, leading to the exclusion of approximately 14--30\% of experimental instances from the paired comparison, as shown in \Cref{tab:paired_bounded_unbounded_summary}. Because these failures occur predominantly in the unbounded formulations, the matched-pair statistics exclude many cases in which bounded models still return finite predictions, thereby providing a conservative estimate of their practical benefit.

\begin{figure}[tb]
\centering
\includegraphics[width=0.95\linewidth]{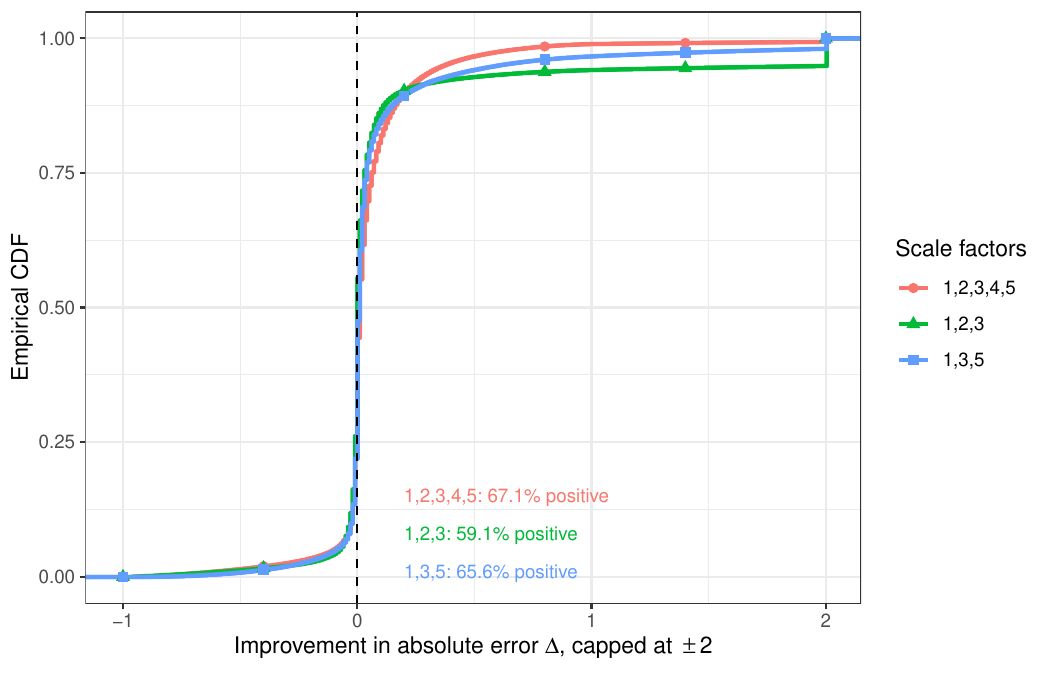}
\caption{Empirical cumulative distribution functions (CDFs) of the improvement in absolute error achieved by the bounded polynomial--exponential extrapolator with asymptote $a=0$ and degree $d=1$ relative to its unbounded counterpart, defined as per \Cref{eq:improvement}. Positive values indicate cases where the bounded model yields a more accurate estimate of the ideal expectation value. Each panel corresponds to a different set of noise scale factors $\lambda$. The annotation in each panel reports the percentage of circuits for which the bounded model achieves lower error than the unbounded model. For visual clarity, improvement values exceeding $\pm 2$ are truncated, and the empirical CDF is computed on these winsorized values to reduce visual clutter.}
\label{fig:sample_ecdf}
\end{figure}

\paragraph{Distribution of improvements}
To clarify the practical implications, we examine the empirical distribution of $\Delta$. \Cref{fig:sample_ecdf} shows the empirical CDFs for the bounded polynomial--exponential model with asymptote $a=0$ and degree $d=1$, selected because it ranks among the lowest in mean error in \Cref{tab:paired_bounded_unbounded_summary} and provides broad matched coverage across configurations. Bounded extrapolation yields lower absolute error in 59--67\% of experiments across the three noise-scale configurations. The magnitude of improvement varies substantially: while the largest degradations approach $-1$ in absolute-error units, improvements exceeding $2$ are occasionally observed.

Taken together, these results show that bounded extrapolation consistently shifts the error distribution toward greater accuracy and, crucially, eliminates catastrophic failures (divergent or non-finite predictions) that arise from unstable unconstrained fits.

\section{Hardware Experiments}
\label{sec:hardware}

While the synthetic benchmark described in \Cref{sec:synthetic} enables controlled, large-scale experimentation, it relies on simulated noise models. To evaluate whether the improvements observed in simulation transfer to real devices, we perform an out-of-distribution validation study using circuits executed on the IBM \texttt{Kingston} quantum processor, part of the Heron r2 processor family~\cite{ibm_processors}. At the time of writing, IBM \texttt{Kingston} exhibits slightly faster execution and lower noise levels than IBM \texttt{Fez}, making it a suitable platform for large-scale hardware experiments. The goal of this experiment is to assess whether the improvements observed for bounded extrapolation models in simulation persist when applied to large circuits on real quantum hardware.

The purpose of this experiment is not to reproduce the large-scale statistical evaluation performed in simulation, but rather to assess whether the qualitative behaviour of bounded extrapolation models persists when applied to substantially larger circuits on real quantum hardware. Due to runtime constraints and queue availability on shared quantum processors, only a limited number of circuits can be executed. Nevertheless, these experiments allow us to examine the stability and accuracy of bounded extrapolation models under realistic device noise and for circuit sizes well beyond those in the synthetic benchmark.

\subsection{Benchmark Circuits}

For the hardware experiments, we focus on two algorithmic circuits from the MQTBench benchmark suite v.2.2.1~\cite{quetschlich2023mqt}: preparation of the Greenberger--Horne--Zeilinger (GHZ) state~\cite{greenberger1989going} and preparation of the W-state~\cite{dur2000three}.\footnote{We also experimented with deeper circuits, such as the Quantum Fourier Transform~\cite{nielsen2010quantum}. However, preliminary experiments indicated that the substantially greater circuit depth caused accumulated hardware noise to dominate the signal, rendering the resulting measurements unsuitable for reliable extrapolation. Consequently, these circuits were excluded from the final hardware study.
}

These two circuits were selected because they yield analytically known global Pauli expectation values. For each circuit, we evaluate both $Z$-type and $X$-type Pauli strings, specifically observables of the form $Z^{\otimes n} = Z_1 Z_2 \cdots Z_n$ and $X^{\otimes n} = X_1 X_2 \cdots X_n$. The corresponding ideal expectation values are given in \Cref{tab:circuit_stats}.\footnote{Ideal values were derived analytically, validated by noiseless simulations on 2--10 qubits, and then extended to larger circuit sizes.} Across the considered circuits and observables, the ideal values lie in $\{-1, 0, 1\}$, allowing us to examine the behaviour of extrapolation models both near the boundaries and in the interior of the physically valid interval $[-1,1]$. This provides a simple way to assess the quality and stability of the extrapolated estimates.

\begin{table}[tb]
\centering
\caption{Circuit depth and gate count for the base circuits ($\lambda=1$).}
\label{tab:circuit_stats}
\resizebox{\columnwidth}{!}{%
\begin{tabular}{@{}lrrrrrr@{}}
\toprule
Circuit & Qubits & Pauli  &  $E^{\mathrm{ideal}}$ & Depth & Overall  & Control (\textsc{cz}) \\
 & ($n$)  &  string &  &  &  gate count & gate count\\

\midrule
GHZ & 30 & $X^{\otimes n}$ & 1 &  94 & 297 & 29\\
GHZ & 30 & $Z^{\otimes n}$ & 1 &  91 & 207 & 29\\
GHZ & 40 & $X^{\otimes n}$ & 1 &  124 & 397 & 39\\
GHZ & 40 & $Z^{\otimes n}$ & 1 &  121 & 277 & 39\\
GHZ & 50 & $X^{\otimes n}$ & 1 &  154 & 497 & 49\\
GHZ & 50 & $Z^{\otimes n}$ & 1 &  151 & 347  & 49\\
\midrule
W-state & 30 & $X^{\otimes n}$ & 0 & 156 & 528 & 58\\
W-state & 30 & $Z^{\otimes n}$ & -1 & 153 & 438  & 58\\
W-state & 40 & $X^{\otimes n}$ & 0 & 206 & 708 & 78\\
W-state & 40 & $Z^{\otimes n}$ & -1 &  203 & 588  & 78\\
W-state & 50 & $X^{\otimes n}$ & 0 &  256 & 888 & 98\\
W-state & 50 & $Z^{\otimes n}$ & -1 & 253 & 738  & 98\\
\bottomrule
\end{tabular}
}
\end{table}

Unlike the synthetic benchmark, which focuses on circuits small enough to be simulated classically, the hardware experiments target substantially larger circuits with qubit counts $n \in \{30,40,50\}$. This range spans the boundary of classical tractability ($30$ qubits) and extends into regimes where full statevector simulation is impractical ($40$ and $50$ qubits), enabling the study of ZNE both near and beyond the limits of efficient classical simulation.

Wider circuits were not considered because circuit depth also increases with width, leading to greater noise accumulation and less reliable extrapolation. Moreover, although the IBM \texttt{Kingston} processor provides 156 physical qubits, calibration quality is non-uniform across the device. Limiting the study to at most 50 qubits allows the transpiler to map circuits onto well-connected subsets of higher-quality qubits and couplers, reducing the risk that hardware errors overwhelm the signal.

\subsection{Hardware Execution}
Before transpilation, we apply a calibration-based filtering step to the device coupling graph. Qubits with high readout or gate errors are excluded before layout selection. Although this filtering reduces the available placement options and can lead to circuits with greater depth or gate count, preliminary tests showed that including these qubits produced measurements dominated by noise, obscuring the observable signal. This choice, therefore, reflects a trade-off between circuit compactness and measurement reliability.

Specifically, physical qubits with readout error exceeding $0.05$ are excluded, and couplings whose calibrated two-qubit gate error exceeds $0.1$ for either the native CZ interaction or the minimum available two-qubit gate are removed. The resulting restricted coupling graph is then supplied to the transpiler, so that circuit placement and routing are performed only on comparatively reliable qubits and couplers. This approach does not necessarily yield the globally optimal transpilation for a given circuit; however, because the primary goal of this study is to assess the behaviour of the mitigation protocol rather than to optimize circuit compilation, the procedure provides a consistent and controlled experimental setting. For the calibration snapshot used in this study, this filtering excluded 12 to 17 qubits (depending on the run conducted on different days) due to readout error. It removed two couplings, leaving 296 edges in the effective coupling graph.

Each circuit is subsequently transpiled 100 times at optimization level~3 using this restricted coupling map, and the best result is selected based on circuit depth (breaking ties by gate count). The resulting layout is then fixed and reused for all folded variants of the circuit, ensuring that differences in measurement results arise from noise scaling rather than changes in qubit mapping. Representative layout statistics are deferred to Appendix~\ref{app:hardware_layout_examples}.

After the layout is fixed, noise amplification is implemented via random gate folding using the Mitiq \texttt{fold\_gates\_at\_random} procedure. Because large noise scale factors rapidly increase circuit depth and hardware cost and may cause noise to overwhelm the signal, we restrict the scale factors to $\lambda \in \{1, 1.3, 1.6 \}$. These fractional values align with those used in earlier experimental ZNE studies and reflect prior observations that moderate noise amplification can improve extrapolation stability compared to very large scale factors~\cite{kim2023scalable}.

Pauli observables are measured by applying the appropriate basis rotations before measuring in the computational basis. Circuit execution uses the Qiskit Runtime \texttt{SamplerV2} primitive~\cite{javadi2024quantum}, with each circuit executed using 10{,}000 measurement shots. No built-in runtime error mitigation or resilience options are enabled, ensuring that the comparison isolates the behaviour of ZNE extrapolation models.

\subsection{Experimental Protocol and Evaluation}

Due to the substantial hardware cost of executing large circuits, the number of repetitions per circuit configuration is limited. For each circuit and noise-scale configuration, we perform measurements using five independent folding seeds, each repeated three times, resulting in a total of fifteen executions per configuration.

This limited number of repetitions reflects practical constraints on available hardware, runtime, and shot budgets. As a result, the collected dataset is too small to support statistically rigorous paired hypothesis testing, such as the Wilcoxon signed-rank test used in the synthetic benchmark.

Instead, the hardware study focuses on qualitative comparisons of extrapolation behaviour and mitigation outcomes across the different model families. We evaluate the same extrapolation models used in the synthetic benchmark, including conventional (unbounded) models and the physically bounded formulations introduced in \Cref{sec:orig_models,sec:bounded}. Mitigation accuracy is measured using \Cref{eq:mae,eq:mse}.

\subsection{Results}

\subsubsection{Observed Noise-Scale Behaviour}\label{sec:observed_noise_behaviour}

\begin{figure}[tb]
\centering
\includegraphics[width=\linewidth]{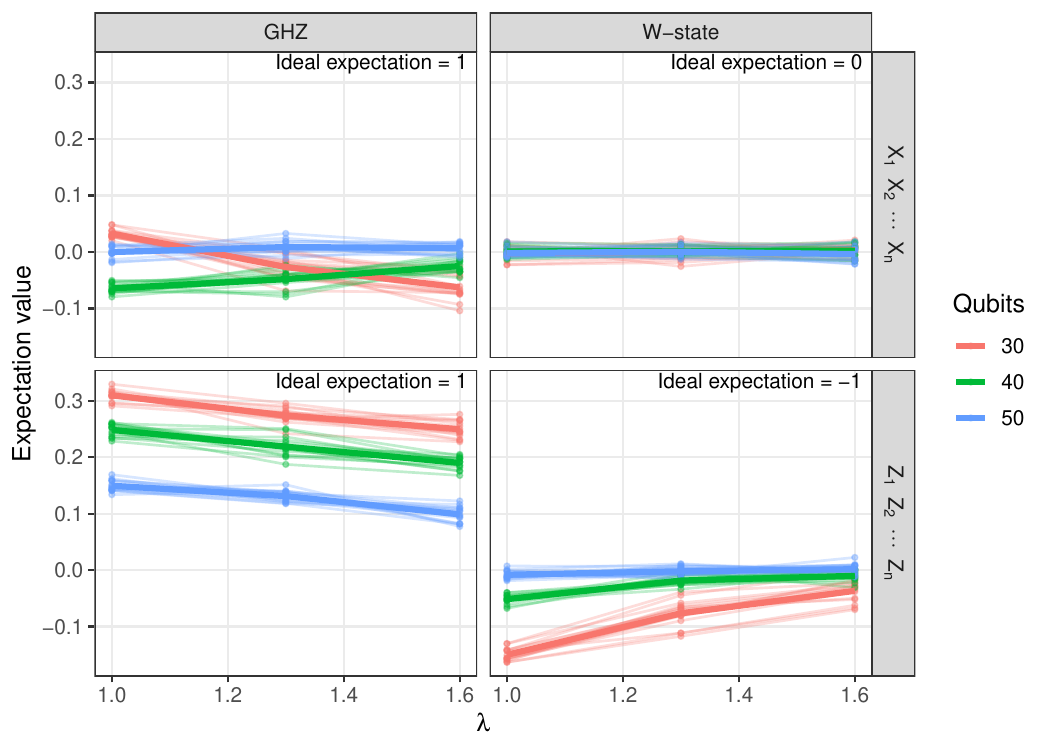}
\caption{Expectation values as a function of the noise scale factor~$\lambda$ for GHZ and W-state circuits. Thin lines show individual repetitions (15 per group), while thick lines indicate the mean expectation value for each configuration.}
\label{fig:zne_summary}
\end{figure}

\Cref{fig:zne_summary} shows the expectation values as a function of the noise scale factor~$\lambda$ for the GHZ and W-state circuits. As the number of qubits increases, extrapolation becomes more challenging, and the observed expectation values cluster closer to zero. This effect is most pronounced for the 50-qubit W-state circuit with the $Z^{\otimes 50}$ observable, which also has the largest width and depth among the considered benchmarks (see \Cref{tab:circuit_stats}). As circuit depth and gate count increase, accumulated noise grows, degrading the signal available for extrapolation~\cite{giurgica2020digital}.

The dependence on~$\lambda$ is also less regular than expected. In several cases, curves that would ideally be concave appear convex, and vice versa. These inconsistencies complicate extrapolation, as the empirical trends may not align with standard model assumptions.

For the GHZ-state circuit measured with the $X^{\otimes n}$ observable, the ideal expectation is $+1$, yet for larger circuits the observed hardware values are often near zero or even negative. We first verified the circuit's correctness for all sizes (up to 50 qubits) using Clifford simulation. To assess expected noise behaviour, we used Qiskit Aer with device-derived noise models. For small instances (e.g., 10 qubits), these simulations agree with hardware, yielding values near $0.68$ at $\lambda=1$. 

However, this agreement does not hold at larger scales. For 30-qubit circuits, Aer simulations still yield substantially higher positive values, $\hat E(1) \approx 0.4$, whereas the corresponding hardware values remain below $0.1$. This gap indicates that the discrepancy is not due to circuit construction but rather to noise processes or device effects not fully captured by the simulation model, which appears optimistic relative to the physical device. A plausible contributing factor is that $X^{\otimes n}$ measurements are more fragile than $Z^{\otimes n}$ measurements: they require basis rotations on all qubits and depend on preserving coherent multi-qubit correlations across the register. As circuit width increases, accumulated gate, decoherence, and readout errors may therefore suppress the $X$-string signal more strongly than predicted by simplified noise models. Although the precise origin of this discrepancy remains open, the hardware results reflect the practically relevant regime encountered by ZNE users.

A related but distinct issue arises for the W-state with the $X^{\otimes n}$ observable, whose ideal expectation is $0$. In this setting, measured values near zero are not necessarily informative, since noise drives expectation values toward zero. Such observables, therefore, provide a weaker diagnostic of extrapolation quality, because apparent agreement with the ground truth may occur trivially. Nevertheless, we include these results for completeness, so that both $X$- and $Z$-type observables are represented for the GHZ and W-state circuits.

\subsubsection{Quantitative Comparison of Extrapolation Models}

The results in \Cref{tab:paired-bounded-unbounded-real-ghz,tab:paired-bounded-unbounded-real-w-state} broadly align with the synthetic benchmark while highlighting hardware-specific effects. Each configuration allows up to $k=15$ matched instances, but many models (particularly those with unconstrained asymptotes) achieve substantially lower coverage (often $k=1$--$4$). As a result, model performance must be interpreted jointly in terms of error, variance, and effective sample size: low-error results at very small $k$ are less informative than slightly higher-error results with near-complete coverage.

Across circuit widths, the $n=30$ setting is generally the most informative for observables with nonzero ideal values, as it exhibits the least signal degradation. For GHZ and W-state circuits with $Z^{\otimes n}$ observables, first-degree polynomial MAE increases with width (e.g., from $\approx 0.59$ to $0.76$ for GHZ-$Z$, and from $\approx 0.67$ to $0.98$ for W-$Z$ between $n=30$ and $n=50$), with full coverage throughout. For GHZ with $X^{\otimes n}$, the $n=30$ case remains challenging but retains more signal than the near-collapse observed at $n=40$ and $n=50$.

Polynomial models show little difference between bounded and unbounded variants: mean errors are nearly identical, and unphysical extrapolations are rare. However, degree-$2$ polynomials are less stable than degree-$1$, exhibiting consistently higher variance despite similar mean errors. Consequently, $d=1$ provides the most reliable high-coverage behaviour among polynomial models.

As discussed in \Cref{sec:observed_noise_behaviour}, for larger widths ($n=40,50$), hardware noise overwhelms the GHZ $X$-string signal, driving the MAE of all high-coverage polynomial models near $1$. Because this occurs at full coverage ($k=15$), it reflects a limitation of the hardware regime rather than of the extrapolation method.

Exponential and polynomial--exponential models follow a different pattern. With a fixed asymptote ($a=0$), both bounded and unbounded variants often achieve high coverage. However, full coverage does not guarantee meaningful extrapolation: half of the 12 unbounded configurations yield extremely large errors despite numerical convergence, whereas the corresponding bounded models remain finite and physically plausible. Thus, bounding primarily improves robustness rather than convergence alone.

When the asymptote is unconstrained, coverage degrades substantially. Unbounded exponential and polynomial--exponential models agree on only a small subset of instances; across 24 such configurations, the unbounded variant attains higher finite coverage in only 5 cases. In these regimes, bounded formulations typically increase the fraction of usable (finite) extrapolations.

Low-coverage results should therefore be interpreted cautiously. Some asymptote-free bounded models report the lowest MAE or MSE across several benchmarks. However, these results are typically based on only $k=1$--$4$ matched instances (at most $k=7$), providing weaker evidence than the consistently observed high-coverage results from degree-$1$ polynomial models.

The W-state with the $X^{\otimes n}$ observable requires separate interpretation because its ideal value is $0$. In this case, low error can arise trivially from noise-induced decay toward zero, making it a weak diagnostic of extrapolation quality. Accordingly, although degree-$1$ polynomial models achieve low error with full coverage, these results are not directly comparable to those for GHZ-$X$, GHZ-$Z$, or W-state-$Z$, where drift toward zero reflects signal loss.

Overall, two conclusions emerge. First, reliable comparisons require high matched-pair coverage ($k \approx 15$). Second, in regimes where optimization is unstable, bounded formulations offer a clear robustness advantage for exponential-family models by increasing finite coverage and avoiding extreme extrapolations. In contrast, many of the strongest asymptote-free results occur at very small $k$ and should be treated as indicative rather than conclusive.

Given the limited number of hardware experiments and variability in matched-pair counts, these findings should be interpreted as suggestive rather than statistically definitive.

\section{Discussion}
\label{sec:discussion}

The results presented in \Cref{sec:synthetic,sec:hardware} highlight several observations about the behaviour of ZNE in practical settings.

\subsection{Importance of Physical Constraints}

A central finding of this study is that imposing physical constraints during model fitting improves the robustness of zero-noise extrapolation. Conventional ZNE models are typically fitted with unconstrained regression, which can yield extrapolated zero-noise predictions outside the physically valid interval $[-1,1]$. For observables with $\pm 1$ eigenvalues, such predictions are not physically meaningful and can substantially degrade mitigation quality.

The bounded formulations introduced in \Cref{sec:bounded} address this issue by incorporating the constraint directly into the optimization problem. Unlike post hoc clipping, this approach allows the physical bounds to shape the fitted curve. In the synthetic benchmark, this change eliminates unphysical predictions and improves the stability and mean accuracy of exponential-family models across a broad range of circuits and noise regimes.

At the same time, the benefit is model-dependent. For polynomial extrapolators, bounded and unbounded variants are usually very similar because polynomial fits in our benchmark rarely yield strongly unphysical zero-noise estimates. Small differences do arise in some cases, both in simulation and on hardware, but these effects are modest compared with the improvements observed for exponential and polynomial--exponential models.

\subsection{Practical Implications for Model Choice}

Beyond the general benefit of enforcing physical bounds, the experiments also provide guidance on which model families are most practical. Across the synthetic benchmark, the strongest overall performance is achieved by low-capacity exponential-family models with simple, physically motivated assumptions, particularly the polynomial--exponential model with degree $d=1$ and a fixed asymptote $a=0$.

This observation is important for two reasons. First, it shows that adding flexibility does not necessarily improve extrapolation quality when only a small number of noise-scale points are available. Second, it suggests that incorporating even limited prior knowledge about the observable can substantially improve robustness. For traceless Pauli observables, fixing the asymptote to $a=0$ reduces the number of free parameters and stabilizes both bounded and unbounded fits, with the bounded versions remaining the most reliable overall.

The hardware study reinforces the importance of jointly evaluating models for accuracy and coverage. Some models achieve good error metrics only on a subset of runs, while others produce usable estimates much more consistently. In practical deployments, this distinction matters: a model that occasionally achieves low error but often fails to converge or yields pathological extrapolations may be less useful than a slightly less accurate model that behaves reliably across executions. From this perspective, bounded formulations provide a clear practical advantage.

\subsection{Relation Between Simulation and Hardware Results}

The hardware experiments broadly support the trends observed in simulation but also reveal additional complications that arise on real devices. As circuits become wider and deeper, measured expectation values tend to drift toward zero, often more rapidly than simulations predict (see \Cref{sec:observed_noise_behaviour}), reducing the signal available for extrapolation. In addition, the dependence on the noise scale factor can become irregular. In several cases, empirical curves exhibit shapes that do not align with the simple monotone or smoothly decaying trends assumed by standard extrapolation models.

\subsection{Practical Deployment Considerations}

An important design goal of this work is to preserve compatibility with standard ZNE workflows. The proposed bounded extrapolation models retain the same functional forms as commonly used polynomial, exponential, and polynomial--exponential extrapolators, and can therefore be integrated into existing software stacks with minimal disruption. In practice, adopting the bounded variants requires only replacing the unconstrained fitting routine with a bound-constrained optimization step; the remainder of the ZNE pipeline is unchanged.

The proposed approach also does not require global training across circuits or hardware devices. Each ZNE instance is processed independently using only the measured expectation values at different noise levels. This makes bounded extrapolation attractive in practical settings where large calibration datasets are unavailable or expensive to obtain.

These properties make the approach complementary to machine-learning-based quantum error mitigation methods that rely on supervised calibration data~\cite{liao2024machine}. More reliable extrapolated estimates may improve the datasets used to train such models, suggesting that bounded extrapolation could also play a useful supporting role in learning-based mitigation pipelines.

\subsection{Limitations}

This study has several limitations. First, the synthetic benchmark relies on simulated device noise and a reduced-device simulation strategy in which only the qubits used by the circuit are modelled. As discussed in \Cref{sec:synthetic}, this makes the simulated noise somewhat optimistic relative to full hardware behaviour.

Second, the hardware evaluation is necessarily limited in scale. Because of runtime and queue constraints, we evaluate only a small number of circuit families, observables, and qubit widths, and use only three relatively mild noise scale factors, $\lambda \in \{1,1.3,1.6\}$. These choices are appropriate for an initial hardware validation, but they do not support statistically strong conclusions of the kind possible in the synthetic benchmark.

Third, the hardware evaluation is inherently subject to variability that is difficult to control and reproduce. Device calibration, noise characteristics, and qubit quality can change from day to day, affecting circuit placement, error rates, and observed expectation values. In contrast, simulations are fully controlled and repeatable, but may provide an optimistic approximation of device behaviour because they do not capture all sources of noise and variability. These factors complicate direct comparison between simulated and real-device results and underscore the importance of hardware validation.

More broadly, the present study focuses on observables with known physical bounds and on a relatively small set of extrapolation families. Extending the analysis to broader circuit classes, additional hardware platforms, and other constrained extrapolation formulations would help clarify the generality of the conclusions.

\section{Conclusion}
\label{sec:conclusion}

This work investigated whether enforcing physical constraints can improve the reliability of zero-noise extrapolation for quantum observables with bounded expectation values. We introduced physically bounded variants of polynomial, exponential, and polynomial--exponential extrapolation models by explicitly parameterizing the zero-noise estimate and constraining it during optimization.

Across a large-scale synthetic benchmark comprising 180{,}000 circuits and $\approx$ 3.6 million ZNE experiments, bounded extrapolation substantially reduced unphysical predictions and improved the stability and accuracy of exponential-family models. In contrast, polynomial models showed little difference between bounded and unbounded formulations, indicating that the benefits of bounding are most pronounced for model families that are more prone to unstable fits. Among the evaluated methods, simple bounded exponential-family models with physically motivated asymptotic assumptions provided the best overall balance of robustness and accuracy.

Preliminary experiments on IBM quantum hardware further support these findings: bounded models remain more stable, avoid pathological extrapolations, and maintain competitive accuracy. At the same time, the hardware results highlight important practical limitations. In particular, noise effects can be more severe and less regular than simulations suggest, and device variability can affect reproducibility across runs. These observations underscore that evaluating error-mitigation methods under real-device conditions is essential to understanding their practical utility.

Overall, enforcing physical constraints during extrapolation provides a simple way to improve the robustness of ZNE in realistic settings without altering the surrounding workflow. Future work should extend hardware validation to broader classes of circuits and devices, and explore combinations with other mitigation and learning-based approaches.

\section*{Acknowledgments}
This work was partially supported by the Natural Sciences and Engineering Research Council of Canada (grant \# RGPIN-2022-03886). The authors thank the Digital Research Alliance of Canada and IBM for providing computational resources.

The authors also used large language models for language editing, proofreading, and assistance in generating and refining portions of the code used in this study. All code was subsequently reviewed by the authors.

\bibliographystyle{IEEEtran}
\bibliography{references}

\clearpage
\appendices

\section{Representative Hardware Layout Statistics}
\label{app:hardware_layout_examples}

For completeness, we report representative statistics of the physical qubits and couplings selected after calibration-based filtering and transpilation.

For the 50-qubit circuits used in the experiments, the selected layouts typically involve physical qubits whose readout errors range from approximately $2.6 \times 10^{-3}$ to $4.1 \times 10^{-2}$, with 90\% of qubits exhibiting readout errors below $2.6 \times 10^{-2}$. The two-qubit interactions used by the transpiled circuits correspond to calibrated CZ error rates between $8.5\times10^{-4}$ and $1.6\times10^{-2}$, with 90\% of couplings exhibiting error rates below $5.7 \times 10^{-3}$. Each interaction is used only a small number of times within the circuits (typically twice per edge), limiting the accumulation of two-qubit gate errors along any individual coupling.

\section{Extended Quantitative Results and Model Comparisons}
\label{app:model_comparison}

In this appendix, we present the full quantitative results underlying the analysis in \Cref{sec:synthetic,sec:hardware}. The tables in this appendix provide the summaries of mitigation performance across all evaluated extrapolation models and experimental configurations. They report aggregate statistics for matched bounded and unbounded model pairs, including the number of matched instances, coverage percentages, and error metrics. 

\Cref{tab:paired_bounded_unbounded_summary} summarizes results for the synthetic benchmark, while \Cref{tab:paired-bounded-unbounded-real-ghz,tab:paired-bounded-unbounded-real-w-state} report results obtained on real quantum hardware for the GHZ and W-state algorithms, respectively. For each configuration, we report the mean absolute error and the mean squared error, along with their associated standard deviations. Standard deviations comparable in magnitude to the mean error should be interpreted primarily as reflecting heterogeneity in circuit difficulty, noise regime, and ideal expectation values across the benchmark, rather than instability of the bounded extrapolation procedure. In addition to error magnitudes, the tables explicitly capture effective sample size through the matched-pair count $k$ and the corresponding coverage statistics.

\begin{landscape}
\begin{table}[t]
\centering
\caption{Aggregate mitigation error across matched bounded/unbounded model pairs for the synthetic dataset. Each cell reports bounded / unbounded values. Top line: matched-pair count $k$, with bracketed coverage percentages (bounded finite / unbounded finite / matched pairs), each normalized by an estimated maximum number of observations (10 $\times$ 179,979 distinct circuit paths $=$ 1,799,790). Rows may remain unmatched due to optimization failure or finite-value filtering. Bold red values indicate the smallest bounded MAE/MSE in that column; bold blue values indicate the smallest unbounded MAE/MSE in that column. Second line: MAE $\pm$ standard deviation. Third line: MSE $\pm$ standard deviation.}
\label{tab:paired_bounded_unbounded_summary}
\setlength{\tabcolsep}{3pt}
\resizebox{1.31\textwidth}{!}{%
\begin{tabular}{@{}p{2.9cm}cccccc@{}}
\toprule
 & \multicolumn{3}{c}{IBM Algiers (simulator)} & \multicolumn{3}{c}{IBM Fez  (simulator)} \\
\cmidrule(lr){2-4} \cmidrule(lr){5-7}
Model family (hyperparameters) & $\lambda = \{1,2,3\}$ & $\lambda = \{1,2,3,4,5\}$ & $\lambda = \{1,3,5\}$ & $\lambda = \{1,2,3\}$ & $\lambda = \{1,2,3,4,5\}$ & $\lambda = \{1,3,5\}$ \\
\midrule
Polynomial ($d=1$) & \makecell[l]{$k$=1,799,783 (100.00 \%/ 100.00 \%/ 100.00\%) \\ MAE: 2.5E-1 $\pm$ 2.2E-1 / 2.5E-1 $\pm$ 2.2E-1 \\ MSE: 1.1E-1 $\pm$ 1.7E-1 / 1.1E-1 $\pm$ 1.7E-1} & \makecell[l]{$k$=1,799,783 (100.00 \%/ 100.00 \%/ 100.00\%) \\ MAE: 2.9E-1 $\pm$ 2.4E-1 / 2.9E-1 $\pm$ 2.4E-1 \\ MSE: 1.4E-1 $\pm$ 2.0E-1 / 1.4E-1 $\pm$ 2.0E-1} & \makecell[l]{$k$=1,799,783 (100.00 \%/ 100.00 \%/ 100.00\%) \\ MAE: 2.8E-1 $\pm$ 2.3E-1 / 2.8E-1 $\pm$ 2.3E-1 \\ MSE: 1.3E-1 $\pm$ 1.9E-1 / 1.3E-1 $\pm$ 1.9E-1} & \makecell[l]{$k$=1,799,786 (100.00 \%/ 100.00 \%/ 100.00\%) \\ MAE: 3.0E-1 $\pm$ 2.4E-1 / 3.0E-1 $\pm$ 2.4E-1 \\ MSE: 1.5E-1 $\pm$ 2.0E-1 / 1.5E-1 $\pm$ 2.0E-1} & \makecell[l]{$k$=1,799,786 (100.00 \%/ 100.00 \%/ 100.00\%) \\ MAE: 3.4E-1 $\pm$ 2.5E-1 / 3.4E-1 $\pm$ 2.5E-1 \\ MSE: 1.8E-1 $\pm$ 2.2E-1 / 1.8E-1 $\pm$ 2.2E-1} & \makecell[l]{$k$=1,799,786 (100.00 \%/ 100.00 \%/ 100.00\%) \\ MAE: 3.2E-1 $\pm$ 2.5E-1 / 3.2E-1 $\pm$ 2.5E-1 \\ MSE: 1.7E-1 $\pm$ 2.1E-1 / 1.7E-1 $\pm$ 2.1E-1} \\
\midrule
Polynomial ($d=2$) & \makecell[l]{$k$=1,799,525 (99.99 \%/ 100.00 \%/ 99.99\%) \\ MAE: 1.9E-1 $\pm$ 1.9E-1 / \textcolor{blue}{\textbf{1.9E-1}} $\pm$ 1.9E-1 \\ MSE: 7.2E-2 $\pm$ 1.4E-1 / \textcolor{blue}{\textbf{7.2E-2}} $\pm$ 1.4E-1} & \makecell[l]{$k$=1,799,777 (100.00 \%/ 100.00 \%/ 100.00\%) \\ MAE: 2.2E-1 $\pm$ 2.1E-1 / 2.2E-1 $\pm$ 2.1E-1 \\ MSE: 9.3E-2 $\pm$ 1.6E-1 / 9.3E-2 $\pm$ 1.6E-1} & \makecell[l]{$k$=1,799,780 (100.00 \%/ 100.00 \%/ 100.00\%) \\ MAE: 2.1E-1 $\pm$ 2.1E-1 / \textcolor{blue}{\textbf{2.1E-1}} $\pm$ 2.1E-1 \\ MSE: 9.1E-2 $\pm$ 1.6E-1 / \textcolor{blue}{\textbf{9.1E-2}} $\pm$ 1.6E-1} & \makecell[l]{$k$=1,799,638 (99.99 \%/ 100.00 \%/ 99.99\%) \\ MAE: 2.3E-1 $\pm$ 2.2E-1 / \textcolor{blue}{\textbf{2.3E-1}} $\pm$ 2.2E-1 \\ MSE: 9.9E-2 $\pm$ 1.7E-1 / \textcolor{blue}{\textbf{9.9E-2}} $\pm$ 1.7E-1} & \makecell[l]{$k$=1,799,786 (100.00 \%/ 100.00 \%/ 100.00\%) \\ MAE: 2.6E-1 $\pm$ 2.3E-1 / 2.6E-1 $\pm$ 2.3E-1 \\ MSE: 1.2E-1 $\pm$ 1.9E-1 / 1.2E-1 $\pm$ 1.9E-1} & \makecell[l]{$k$=1,799,786 (100.00 \%/ 100.00 \%/ 100.00\%) \\ MAE: 2.6E-1 $\pm$ 2.3E-1 / \textcolor{blue}{\textbf{2.6E-1}} $\pm$ 2.3E-1 \\ MSE: 1.2E-1 $\pm$ 1.9E-1 / \textcolor{blue}{\textbf{1.2E-1}} $\pm$ 1.9E-1} \\
\midrule
Polynomial ($d=3$) & \makecell[l]{Not enough data points for this degree} & \makecell[l]{$k$=1,625,711 (90.33 \%/ 100.00 \%/ 90.33\%) \\ MAE: 1.8E-1 $\pm$ 1.9E-1 / \textcolor{blue}{\textbf{1.8E-1}} $\pm$ 1.9E-1 \\ MSE: 6.6E-2 $\pm$ 1.3E-1 / \textcolor{blue}{\textbf{6.6E-2}} $\pm$ 1.3E-1} & \makecell[l]{Not enough data points for this degree} & \makecell[l]{$k$=1,626,355 (90.36 \%/ 100.00 \%/ 90.36\%) \\ MAE: 2.2E-1 $\pm$ 2.1E-1 / \textcolor{blue}{\textbf{2.2E-1}} $\pm$ 2.1E-1 \\ MSE: 9.3E-2 $\pm$ 1.6E-1 / \textcolor{blue}{\textbf{9.3E-2}} $\pm$ 1.6E-1} & \makecell[l]{Not enough data points for this degree} & \makecell[l]{Not enough data points for this degree} \\
\midrule
Exponential ($a=0$) & \makecell[l]{$k$=1,799,737 (100.00 \%/ 100.00 \%/ 100.00\%) \\ MAE: 1.4E-1 $\pm$ 1.9E-1 / 1.2E+0 $\pm$ 8.6E+0 \\ MSE: 5.5E-2 $\pm$ 1.5E-1 / 7.6E+1 $\pm$ 5.2E+3} & \makecell[l]{$k$=1,799,736 (100.00 \%/ 100.00 \%/ 100.00\%) \\ MAE: 1.4E-1 $\pm$ 1.8E-1 / 2.6E-1 $\pm$ 2.1E+1 \\ MSE: 5.3E-2 $\pm$ 1.4E-1 / 4.6E+2 $\pm$ 3.2E+5} & \makecell[l]{$k$=1,799,722 (100.00 \%/ 100.00 \%/ 100.00\%) \\ MAE: 1.5E-1 $\pm$ 1.9E-1 / 2.8E-1 $\pm$ 7.3E-1 \\ MSE: 6.1E-2 $\pm$ 1.4E-1 / 6.1E-1 $\pm$ 7.0E+0} & \makecell[l]{$k$=1,799,481 (99.98 \%/ 100.00 \%/ 99.98\%) \\ MAE: 1.7E-1 $\pm$ 2.1E-1 / 1.5E+0 $\pm$ 8.9E+0 \\ MSE: 7.6E-2 $\pm$ 1.8E-1 / 8.1E+1 $\pm$ 1.7E+3} & \makecell[l]{$k$=1,799,460 (99.98 \%/ 100.00 \%/ 99.98\%) \\ MAE: 1.7E-1 $\pm$ 2.1E-1 / 3.8E-1 $\pm$ 5.1E+1 \\ MSE: 7.4E-2 $\pm$ 1.7E-1 / 2.6E+3 $\pm$ 1.8E+6} & \makecell[l]{$k$=1,799,432 (99.98 \%/ 100.00 \%/ 99.98\%) \\ MAE: 1.9E-1 $\pm$ 2.1E-1 / 3.2E-1 $\pm$ 6.8E-1 \\ MSE: 8.1E-2 $\pm$ 1.7E-1 / 5.6E-1 $\pm$ 5.9E+0} \\
\midrule
Exponential ($a$ is unconstrained) & \makecell[l]{$k$=1,296,496 (99.97 \%/ 72.07 \%/ 72.04\%) \\ MAE: 1.6E-1 $\pm$ 1.7E-1 / 3.7E+1 $\pm$ 3.6E+2 \\ MSE: 5.2E-2 $\pm$ 1.1E-1 / 1.3E+5 $\pm$ 1.9E+6} & \makecell[l]{$k$=1,551,721 (99.99 \%/ 86.23 \%/ 86.22\%) \\ MAE: 1.2E-1 $\pm$ 1.5E-1 / 5.1E+1 $\pm$ 4.6E+2 \\ MSE: \textcolor{red}{\textbf{3.9E-2}} $\pm$ 1.1E-1 / 2.1E+5 $\pm$ 3.2E+6} & \makecell[l]{$k$=1,405,985 (99.99 \%/ 78.13 \%/ 78.12\%) \\ MAE: 1.4E-1 $\pm$ 1.9E-1 / 1.5E+0 $\pm$ 6.5E+0 \\ MSE: 5.6E-2 $\pm$ 1.5E-1 / 4.5E+1 $\pm$ 3.2E+2} & \makecell[l]{$k$=1,259,967 (99.96 \%/ 70.04 \%/ 70.01\%) \\ MAE: 1.8E-1 $\pm$ 1.8E-1 / 4.6E+1 $\pm$ 4.0E+2 \\ MSE: 6.6E-2 $\pm$ 1.3E-1 / 1.6E+5 $\pm$ 2.1E+6} & \makecell[l]{$k$=1,513,418 (99.98 \%/ 84.11 \%/ 84.09\%) \\ MAE: \textcolor{red}{\textbf{1.5E-1}} $\pm$ 1.8E-1 / 6.8E+1 $\pm$ 5.3E+2 \\ MSE: \textcolor{red}{\textbf{5.5E-2}} $\pm$ 1.3E-1 / 2.9E+5 $\pm$ 3.8E+6} & \makecell[l]{$k$=1,348,923 (99.98 \%/ 74.96 \%/ 74.95\%) \\ MAE: 1.8E-1 $\pm$ 2.1E-1 / 2.0E+0 $\pm$ 7.5E+0 \\ MSE: 7.8E-2 $\pm$ 1.8E-1 / 6.1E+1 $\pm$ 3.7E+2} \\
\midrule
Poly-exp ($a=0$, $d=1$) & \makecell[l]{$k$=1,797,953 (99.90 \%/ 100.00 \%/ 99.90\%) \\ MAE: \textcolor{red}{\textbf{1.4E-1}} $\pm$ 1.9E-1 / 1.2E+0 $\pm$ 8.6E+0 \\ MSE: 5.4E-2 $\pm$ 1.4E-1 / 7.6E+1 $\pm$ 5.2E+3} & \makecell[l]{$k$=1,798,308 (99.92 \%/ 100.00 \%/ 99.92\%) \\ MAE: 1.4E-1 $\pm$ 1.8E-1 / 2.6E-1 $\pm$ 2.2E+1 \\ MSE: 5.2E-2 $\pm$ 1.4E-1 / 4.6E+2 $\pm$ 3.2E+5} & \makecell[l]{$k$=1,798,118 (99.91 \%/ 100.00 \%/ 99.91\%) \\ MAE: 1.5E-1 $\pm$ 1.9E-1 / 2.8E-1 $\pm$ 7.3E-1 \\ MSE: 6.1E-2 $\pm$ 1.4E-1 / 6.1E-1 $\pm$ 7.0E+0} & \makecell[l]{$k$=1,796,962 (99.84 \%/ 100.00 \%/ 99.84\%) \\ MAE: \textcolor{red}{\textbf{1.7E-1}} $\pm$ 2.1E-1 / 1.5E+0 $\pm$ 8.9E+0 \\ MSE: 7.5E-2 $\pm$ 1.8E-1 / 8.1E+1 $\pm$ 1.7E+3} & \makecell[l]{$k$=1,797,517 (99.87 \%/ 100.00 \%/ 99.87\%) \\ MAE: 1.7E-1 $\pm$ 2.1E-1 / 3.8E-1 $\pm$ 5.1E+1 \\ MSE: 7.3E-2 $\pm$ 1.7E-1 / 2.6E+3 $\pm$ 1.8E+6} & \makecell[l]{$k$=1,797,171 (99.85 \%/ 100.00 \%/ 99.85\%) \\ MAE: 1.9E-1 $\pm$ 2.1E-1 / 3.2E-1 $\pm$ 6.8E-1 \\ MSE: 8.1E-2 $\pm$ 1.7E-1 / 5.6E-1 $\pm$ 5.9E+0} \\
\midrule
Poly-exp ($a=0$, $d=2$) & \makecell[l]{$k$=1,796,810 (99.83 \%/ 100.00 \%/ 99.83\%) \\ MAE: 2.0E-1 $\pm$ 2.1E-1 / 3.4E+10 $\pm$ 5.8E+11 \\ MSE: 8.6E-2 $\pm$ 1.6E-1 / 3.3E+23 $\pm$ 2.4E+25} & \makecell[l]{$k$=1,794,637 (99.71 \%/ 100.00 \%/ 99.71\%) \\ MAE: 1.7E-1 $\pm$ 2.1E-1 / 1.3E+8 $\pm$ 1.3E+11 \\ MSE: 7.2E-2 $\pm$ 1.5E-1 / 1.7E+22 $\pm$ 2.1E+25} & \makecell[l]{$k$=1,797,601 (99.88 \%/ 100.00 \%/ 99.88\%) \\ MAE: 2.0E-1 $\pm$ 2.2E-1 / 2.3E+3 $\pm$ 1.5E+4 \\ MSE: 8.7E-2 $\pm$ 1.6E-1 / 2.3E+8 $\pm$ 3.3E+9} & \makecell[l]{$k$=1,795,896 (99.78 \%/ 100.00 \%/ 99.78\%) \\ MAE: 2.4E-1 $\pm$ 2.3E-1 / 4.6E+10 $\pm$ 6.6E+11 \\ MSE: 1.1E-1 $\pm$ 1.9E-1 / 4.3E+23 $\pm$ 2.0E+25} & \makecell[l]{$k$=1,792,710 (99.61 \%/ 100.00 \%/ 99.61\%) \\ MAE: 2.1E-1 $\pm$ 2.3E-1 / 3.0E+7 $\pm$ 3.3E+10 \\ MSE: 9.8E-2 $\pm$ 1.8E-1 / 1.1E+21 $\pm$ 1.5E+24} & \makecell[l]{$k$=1,796,954 (99.84 \%/ 100.00 \%/ 99.84\%) \\ MAE: 2.4E-1 $\pm$ 2.4E-1 / 2.8E+3 $\pm$ 1.6E+4 \\ MSE: 1.1E-1 $\pm$ 1.9E-1 / 2.7E+8 $\pm$ 3.4E+9} \\
\midrule
Poly-exp ($a=0$, $d=3$) & \makecell[l]{Not enough data points for this degree} & \makecell[l]{$k$=1,794,661 (99.72 \%/ 100.00 \%/ 99.72\%) \\ MAE: 2.6E-1 $\pm$ 2.4E-1 / 8.0E+30 $\pm$ 1.1E+34 \\ MSE: 1.3E-1 $\pm$ 1.9E-1 / 1.2E+68 $\pm$ 1.5E+71} &  \makecell[l]{Not enough data points for this degree} & \makecell[l]{Not enough data points for this degree} & \makecell[l]{$k$=1,792,913 (99.62 \%/ 100.00 \%/ 99.62\%) \\ MAE: 2.9E-1 $\pm$ 2.5E-1 / 7.7E+27 $\pm$ 6.1E+30 \\ MSE: 1.5E-1 $\pm$ 2.1E-1 / 3.8E+61 $\pm$ 3.5E+64} & \makecell[l]{Not enough data points for this degree} \\
\midrule
Poly-exp ($a$ is unconstrained, $d=1$) & \makecell[l]{$k$=1,296,536 (99.97 \%/ 72.07 \%/ 72.04\%) \\ MAE: 1.6E-1 $\pm$ 1.7E-1 / 3.7E+1 $\pm$ 3.6E+2 \\ MSE: \textcolor{red}{\textbf{5.2E-2}} $\pm$ 1.1E-1 / 1.3E+5 $\pm$ 1.9E+6} & \makecell[l]{$k$=1,551,719 (99.99 \%/ 86.23 \%/ 86.22\%) \\ MAE: \textcolor{red}{\textbf{1.2E-1}} $\pm$ 1.5E-1 / 5.1E+1 $\pm$ 4.6E+2 \\ MSE: 3.9E-2 $\pm$ 1.1E-1 / 2.1E+5 $\pm$ 3.2E+6} & \makecell[l]{$k$=1,405,983 (99.99 \%/ 78.13 \%/ 78.12\%) \\ MAE: \textcolor{red}{\textbf{1.4E-1}} $\pm$ 1.7E-1 / 1.5E+0 $\pm$ 6.5E+0 \\ MSE: \textcolor{red}{\textbf{4.8E-2}} $\pm$ 1.2E-1 / 4.5E+1 $\pm$ 3.2E+2} & \makecell[l]{$k$=1,259,984 (99.97 \%/ 70.04 \%/ 70.01\%) \\ MAE: 1.8E-1 $\pm$ 1.8E-1 / 4.6E+1 $\pm$ 4.0E+2 \\ MSE: \textcolor{red}{\textbf{6.6E-2}} $\pm$ 1.3E-1 / 1.6E+5 $\pm$ 2.1E+6} & \makecell[l]{$k$=1,513,420 (99.98 \%/ 84.11 \%/ 84.09\%) \\ MAE: 1.5E-1 $\pm$ 1.8E-1 / 6.8E+1 $\pm$ 5.3E+2 \\ MSE: 5.5E-2 $\pm$ 1.3E-1 / 2.9E+5 $\pm$ 3.8E+6} & \makecell[l]{$k$=1,348,921 (99.98 \%/ 74.96 \%/ 74.95\%) \\ MAE: \textcolor{red}{\textbf{1.7E-1}} $\pm$ 2.0E-1 / 2.0E+0 $\pm$ 7.5E+0 \\ MSE: \textcolor{red}{\textbf{7.0E-2}} $\pm$ 1.5E-1 / 6.1E+1 $\pm$ 3.7E+2} \\
\midrule
Poly-exp ($a$ is unconstrained, $d=2$) & \makecell[l]{Not enough data points for this degree} & \makecell[l]{$k$=1,452,985 (99.88 \%/ 80.77 \%/ 80.73\%) \\ MAE: 1.9E-1 $\pm$ 2.1E-1 / 6.7E+1 $\pm$ 8.8E+3 \\ MSE: 7.8E-2 $\pm$ 1.6E-1 / 7.7E+7 $\pm$ 1.7E+10} & \makecell[l]{Not enough data points for this degree} & \makecell[l]{Not enough data points for this degree} & \makecell[l]{$k$=1,397,119 (99.85 \%/ 77.68 \%/ 77.63\%) \\ MAE: 2.2E-1 $\pm$ 2.3E-1 / 7.7E+1 $\pm$ 1.0E+4 \\ MSE: 1.0E-1 $\pm$ 1.9E-1 / 1.0E+8 $\pm$ 2.8E+10} & \makecell[l]{Not enough data points for this degree} \\
\midrule
Poly-exp ($a$ is unconstrained, $d=3$) & \makecell[l]{Not enough data points for this degree} & \makecell[l]{$k$=1,086,705 (99.87 \%/ 60.45 \%/ 60.38\%) \\ MAE: 2.5E-1 $\pm$ 2.2E-1 / 1.2E+6 $\pm$ 6.7E+8 \\ MSE: 1.1E-1 $\pm$ 1.7E-1 / 4.4E+17 $\pm$ 3.5E+20} & \makecell[l]{Not enough data points for this degree} & \makecell[l]{Not enough data points for this degree} & \makecell[l]{$k$=1,047,604 (99.83 \%/ 58.30 \%/ 58.21\%) \\ MAE: 2.7E-1 $\pm$ 2.3E-1 / 5.4E+6 $\pm$ 2.7E+9 \\ MSE: 1.3E-1 $\pm$ 1.9E-1 / 7.3E+18 $\pm$ 4.7E+21} & \makecell[l]{Not enough data points for this degree} \\
\bottomrule
\end{tabular}
}
\end{table}
\end{landscape}

\begin{landscape}
\begin{table}[t]
\centering
\setlength{\tabcolsep}{3pt}
\caption{Aggregate mitigation error across matched bounded/unbounded model pairs for the GHZ hardware benchmarks. Each cell reports bounded / unbounded values. Top line: matched-pair count $k$, with bracketed coverage percentages (bounded finite / unbounded finite / matched pairs), each normalized by the estimated maximum number of observations (15). Rows may remain unmatched due to optimization failure or finite-value filtering. Bold red values indicate the smallest bounded MAE/MSE within a benchmark column; bold blue values indicate the smallest unbounded MAE/MSE within a benchmark column. Second line: MAE $\pm$ standard deviation. Third line: MSE $\pm$ standard deviation.}
\label{tab:paired-bounded-unbounded-real-ghz}
\resizebox{\linewidth}{!}{%
\begin{tabular}{@{}p{2.9cm}cccccc@{}}
\toprule
 & GHZ, $X^{\otimes 30}$ & GHZ, $X^{\otimes 40}$ & GHZ, $X^{\otimes 50}$ & GHZ, $Z^{\otimes 30}$ & GHZ, $Z^{\otimes 40}$ & GHZ, $Z^{\otimes 50}$  \\
\cmidrule(lr){2-7}
Model family (hyperparameters) & $\lambda = \{1,1.3,1.6\}$ & $\lambda = \{1,1.3,1.6\}$ & $\lambda = \{1,1.3,1.6\}$ & $\lambda = \{1,1.3,1.6\}$ & $\lambda = \{1,1.3,1.6\}$ & $\lambda = \{1,1.3,1.6\}$  \\
\midrule
Polynomial ($d=1$) & \makecell[l]{$k$=15 (100.00\% / 100.00\% / 100.00\%) \\ MAE: 8.1E-1 $\pm$ 4.4E-2 / 8.1E-1 $\pm$ 4.4E-2 \\ MSE: 6.6E-1 $\pm$ 7.0E-2 / 6.6E-1 $\pm$ 7.0E-2} & \makecell[l]{$k$=15 (100.00\% / 100.00\% / 100.00\%) \\ MAE: 1.1E+0 $\pm$ 2.8E-2 / 1.1E+0 $\pm$ 2.8E-2 \\ MSE: 1.3E+0 $\pm$ 6.5E-2 / 1.3E+0 $\pm$ 6.5E-2} & \makecell[l]{$k$=15 (100.00\% / 100.00\% / 100.00\%) \\ MAE: \textcolor{red}{\textbf{1.0E+0}} $\pm$ 2.8E-2 / \textcolor{blue}{\textbf{1.0E+0}} $\pm$ 2.8E-2 \\ MSE: \textcolor{red}{\textbf{1.0E+0}} $\pm$ 5.7E-2 / \textcolor{blue}{\textbf{1.0E+0}} $\pm$ 5.7E-2} & \makecell[l]{$k$=15 (100.00\% / 100.00\% / 100.00\%) \\ MAE: 5.9E-1 $\pm$ 3.6E-2 / 5.9E-1 $\pm$ 3.6E-2 \\ MSE: 3.5E-1 $\pm$ 4.3E-2 / 3.5E-1 $\pm$ 4.3E-2} & \makecell[l]{$k$=15 (100.00\% / 100.00\% / 100.00\%) \\ MAE: 6.5E-1 $\pm$ 3.3E-2 / 6.5E-1 $\pm$ 3.3E-2 \\ MSE: 4.3E-1 $\pm$ 4.3E-2 / 4.3E-1 $\pm$ 4.3E-2} & \makecell[l]{$k$=15 (100.00\% / 100.00\% / 100.00\%) \\ MAE: 7.6E-1 $\pm$ 3.3E-2 / 7.6E-1 $\pm$ 3.3E-2 \\ MSE: 5.8E-1 $\pm$ 5.0E-2 / 5.8E-1 $\pm$ 5.0E-2}  \\
\midrule
Polynomial ($d=2$) & \makecell[l]{$k$=15 (100.00\% / 100.00\% / 100.00\%) \\ MAE: 6.0E-1 $\pm$ 2.8E-1 / 6.0E-1 $\pm$ 2.8E-1 \\ MSE: 4.4E-1 $\pm$ 3.3E-1 / 4.4E-1 $\pm$ 3.3E-1} & \makecell[l]{$k$=15 (100.00\% / 100.00\% / 100.00\%) \\ MAE: \textcolor{red}{\textbf{1.1E+0}} $\pm$ 3.4E-1 / \textcolor{blue}{\textbf{1.1E+0}} $\pm$ 3.4E-1 \\ MSE: \textcolor{red}{\textbf{1.3E+0}} $\pm$ 7.1E-1 / \textcolor{blue}{\textbf{1.3E+0}} $\pm$ 7.1E-1} & \makecell[l]{$k$=15 (100.00\% / 100.00\% / 100.00\%) \\ MAE: 1.1E+0 $\pm$ 2.2E-1 / 1.1E+0 $\pm$ 2.2E-1 \\ MSE: 1.3E+0 $\pm$ 4.7E-1 / 1.3E+0 $\pm$ 4.7E-1} & \makecell[l]{$k$=14 (93.33\% / 100.00\% / 93.33\%) \\ MAE: 5.3E-1 $\pm$ 2.1E-1 / 5.3E-1 $\pm$ 2.1E-1 \\ MSE: 3.2E-1 $\pm$ 2.1E-1 / 3.2E-1 $\pm$ 2.1E-1} & \makecell[l]{$k$=15 (100.00\% / 100.00\% / 100.00\%) \\ MAE: 6.5E-1 $\pm$ 2.9E-1 / 6.5E-1 $\pm$ 2.9E-1 \\ MSE: 5.0E-1 $\pm$ 4.1E-1 / 5.0E-1 $\pm$ 4.1E-1} & \makecell[l]{$k$=15 (100.00\% / 100.00\% / 100.00\%) \\ MAE: 8.9E-1 $\pm$ 2.6E-1 / 8.9E-1 $\pm$ 2.6E-1 \\ MSE: 8.6E-1 $\pm$ 5.1E-1 / 8.6E-1 $\pm$ 5.1E-1} \\
\midrule
Exponential ($a=0$) & \makecell[l]{$k$=15 (100.00\% / 100.00\% / 100.00\%) \\ MAE: 1.0E+0 $\pm$ 1.3E-2 / 3.2E+7 $\pm$ 2.7E+7 \\ MSE: 1.0E+0 $\pm$ 2.7E-2 / 1.7E+15 $\pm$ 3.1E+15} & \makecell[l]{$k$=15 (100.00\% / 100.00\% / 100.00\%) \\ MAE: 1.3E+0 $\pm$ 2.2E-1 / 1.4E+0 $\pm$ 3.1E-1 \\ MSE: 1.8E+0 $\pm$ 6.8E-1 / 1.9E+0 $\pm$ 1.1E+0} & \makecell[l]{$k$=15 (100.00\% / 100.00\% / 100.00\%) \\ MAE: 1.1E+0 $\pm$ 2.7E-1 / 4.8E+5 $\pm$ 1.8E+6 \\ MSE: 1.2E+0 $\pm$ 7.9E-1 / 3.3E+12 $\pm$ 1.3E+13} & \makecell[l]{$k$=15 (100.00\% / 100.00\% / 100.00\%) \\ MAE: 5.5E-1 $\pm$ 5.9E-2 / 5.5E-1 $\pm$ 5.8E-2 \\ MSE: 3.1E-1 $\pm$ 6.6E-2 / 3.1E-1 $\pm$ 6.5E-2} & \makecell[l]{$k$=15 (100.00\% / 100.00\% / 100.00\%) \\ MAE: 6.1E-1 $\pm$ 6.0E-2 / 6.1E-1 $\pm$ 5.8E-2 \\ MSE: 3.7E-1 $\pm$ 7.2E-2 / 3.7E-1 $\pm$ 7.0E-2} & \makecell[l]{$k$=15 (100.00\% / 100.00\% / 100.00\%) \\ MAE: 7.0E-1 $\pm$ 7.1E-2 / 6.9E-1 $\pm$ 7.8E-2 \\ MSE: 4.9E-1 $\pm$ 9.9E-2 / 4.8E-1 $\pm$ 1.1E-1} \\
\midrule
Exponential ($a$ is unconstrained) & \makecell[l]{$k$=3 (53.33\% / 66.67\% / 20.00\%) \\ MAE: 6.0E-1 $\pm$ 3.9E-1 / \textcolor{blue}{\textbf{3.3E-1}} $\pm$ 3.9E-1 \\ MSE: 4.6E-1 $\pm$ 3.8E-1 / \textcolor{blue}{\textbf{2.1E-1}} $\pm$ 3.4E-1} & \makecell[l]{$k$=6 (100.00\% / 40.00\% / 40.00\%) \\ MAE: 1.3E+0 $\pm$ 3.0E-1 / 5.7E+1 $\pm$ 1.4E+2 \\ MSE: 1.7E+0 $\pm$ 9.2E-1 / 1.9E+4 $\pm$ 4.6E+4} & \makecell[l]{$k$=3 (93.33\% / 20.00\% / 20.00\%) \\ MAE: 1.0E+0 $\pm$ 4.1E-2 / 2.1E+0 $\pm$ 1.2E+0 \\ MSE: 1.1E+0 $\pm$ 8.6E-2 / 5.4E+0 $\pm$ 5.6E+0} & \makecell[l]{$k$=6 (80.00\% / 53.33\% / 40.00\%) \\ MAE: 5.9E-1 $\pm$ 4.0E-2 / 4.1E+0 $\pm$ 8.4E+0 \\ MSE: 3.5E-1 $\pm$ 4.8E-2 / 7.6E+1 $\pm$ 1.9E+2} & \makecell[l]{$k$=6 (73.33\% / 66.67\% / 40.00\%) \\ MAE: 6.5E-1 $\pm$ 5.2E-2 / 5.3E-1 $\pm$ 1.1E-1 \\ MSE: 4.2E-1 $\pm$ 6.6E-2 / 2.9E-1 $\pm$ 1.1E-1} & \makecell[l]{$k$=4 (93.33\% / 33.33\% / 26.67\%) \\ MAE: 6.7E-1 $\pm$ 1.5E-1 / 1.5E+1 $\pm$ 2.9E+1 \\ MSE: 4.7E-1 $\pm$ 1.8E-1 / 8.4E+2 $\pm$ 1.7E+3} \\
\midrule
Poly-exp ($a=0$, $d=1$) & \makecell[l]{$k$=13 (86.67\% / 100.00\% / 86.67\%) \\ MAE: 1.0E+0 $\pm$ 5.1E-4 / 3.5E+7 $\pm$ 2.7E+7 \\ MSE: 1.0E+0 $\pm$ 1.0E-3 / 1.9E+15 $\pm$ 3.3E+15} & \makecell[l]{$k$=15 (100.00\% / 100.00\% / 100.00\%) \\ MAE: 1.3E+0 $\pm$ 2.2E-1 / 1.4E+0 $\pm$ 3.1E-1 \\ MSE: 1.8E+0 $\pm$ 6.8E-1 / 1.9E+0 $\pm$ 1.1E+0} & \makecell[l]{$k$=15 (100.00\% / 100.00\% / 100.00\%) \\ MAE: 1.1E+0 $\pm$ 2.7E-1 / 4.8E+5 $\pm$ 1.8E+6 \\ MSE: 1.2E+0 $\pm$ 7.9E-1 / 3.3E+12 $\pm$ 1.3E+13} & \makecell[l]{$k$=15 (100.00\% / 100.00\% / 100.00\%) \\ MAE: 5.5E-1 $\pm$ 5.9E-2 / 5.5E-1 $\pm$ 5.8E-2 \\ MSE: 3.1E-1 $\pm$ 6.6E-2 / 3.1E-1 $\pm$ 6.5E-2} & \makecell[l]{$k$=15 (100.00\% / 100.00\% / 100.00\%) \\ MAE: 6.1E-1 $\pm$ 6.0E-2 / 6.1E-1 $\pm$ 5.8E-2 \\ MSE: 3.7E-1 $\pm$ 7.2E-2 / 3.7E-1 $\pm$ 7.0E-2} & \makecell[l]{$k$=15 (100.00\% / 100.00\% / 100.00\%) \\ MAE: 7.0E-1 $\pm$ 7.1E-2 / 6.9E-1 $\pm$ 7.8E-2 \\ MSE: 4.9E-1 $\pm$ 9.9E-2 / 4.8E-1 $\pm$ 1.1E-1} \\
\midrule
Poly-exp ($a=0$, $d=2$) & \makecell[l]{$k$=15 (100.00\% / 100.00\% / 100.00\%) \\ MAE: 9.9E-1 $\pm$ 4.9E-2 / 1.0E+47 $\pm$ 3.3E+47 \\ MSE: 9.8E-1 $\pm$ 9.0E-2 / 1.1E+95 $\pm$ 4.4E+95} & \makecell[l]{$k$=15 (100.00\% / 100.00\% / 100.00\%) \\ MAE: 1.3E+0 $\pm$ 3.9E-1 / 4.0E+4 $\pm$ 1.5E+5 \\ MSE: 1.7E+0 $\pm$ 1.2E+0 / 2.4E+10 $\pm$ 9.3E+10} & \makecell[l]{$k$=15 (100.00\% / 100.00\% / 100.00\%) \\ MAE: 1.0E+0 $\pm$ 6.1E-2 / 5.3E+69 $\pm$ 1.9E+70 \\ MSE: 1.0E+0 $\pm$ 1.3E-1 / 3.6E+140 $\pm$ 1.4E+141} & \makecell[l]{$k$=11 (73.33\% / 100.00\% / 73.33\%) \\ MAE: 5.6E-1 $\pm$ 2.4E-1 / 5.6E-1 $\pm$ 2.5E-1 \\ MSE: 3.7E-1 $\pm$ 2.2E-1 / 3.7E-1 $\pm$ 2.2E-1} & \makecell[l]{$k$=12 (80.00\% / 100.00\% / 80.00\%) \\ MAE: 7.0E-1 $\pm$ 2.4E-1 / 6.6E-1 $\pm$ 2.5E-1 \\ MSE: 5.4E-1 $\pm$ 2.9E-1 / 4.9E-1 $\pm$ 3.1E-1} & \makecell[l]{$k$=14 (93.33\% / 100.00\% / 93.33\%) \\ MAE: 8.5E-1 $\pm$ 1.6E-1 / 8.3E-1 $\pm$ 2.1E-1 \\ MSE: 7.5E-1 $\pm$ 2.4E-1 / 7.3E-1 $\pm$ 3.0E-1} \\
\midrule
Poly-exp ($a$ is unconstrained, $d=1$) & \makecell[l]{$k$=3 (53.33\% / 66.67\% / 20.00\%) \\ MAE: \textcolor{red}{\textbf{3.3E-1}} $\pm$ 3.9E-1 / \textcolor{blue}{\textbf{3.3E-1}} $\pm$ 3.9E-1 \\ MSE: \textcolor{red}{\textbf{2.1E-1}} $\pm$ 3.4E-1 / \textcolor{blue}{\textbf{2.1E-1}} $\pm$ 3.4E-1} & \makecell[l]{$k$=6 (100.00\% / 40.00\% / 40.00\%) \\ MAE: 1.4E+0 $\pm$ 3.2E-1 / 5.7E+1 $\pm$ 1.4E+2 \\ MSE: 2.1E+0 $\pm$ 1.0E+0 / 1.9E+4 $\pm$ 4.6E+4} & \makecell[l]{$k$=3 (100.00\% / 20.00\% / 20.00\%) \\ MAE: 1.4E+0 $\pm$ 5.4E-1 / 2.1E+0 $\pm$ 1.2E+0 \\ MSE: 2.1E+0 $\pm$ 1.7E+0 / 5.4E+0 $\pm$ 5.6E+0} & \makecell[l]{$k$=4 (66.67\% / 53.33\% / 26.67\%) \\ MAE: \textcolor{red}{\textbf{4.2E-1}} $\pm$ 2.6E-1 / \textcolor{blue}{\textbf{4.3E-1}} $\pm$ 2.6E-1 \\ MSE: \textcolor{red}{\textbf{2.3E-1}} $\pm$ 1.6E-1 / \textcolor{blue}{\textbf{2.3E-1}} $\pm$ 1.6E-1} & \makecell[l]{$k$=7 (80.00\% / 66.67\% / 46.67\%) \\ MAE: \textcolor{red}{\textbf{4.5E-1}} $\pm$ 1.9E-1 / \textcolor{blue}{\textbf{5.1E-1}} $\pm$ 1.1E-1 \\ MSE: \textcolor{red}{\textbf{2.4E-1}} $\pm$ 1.3E-1 / \textcolor{blue}{\textbf{2.7E-1}} $\pm$ 1.1E-1} & \makecell[l]{$k$=3 (86.67\% / 33.33\% / 20.00\%) \\ MAE: \textcolor{red}{\textbf{5.8E-1}} $\pm$ 1.1E-1 / \textcolor{blue}{\textbf{5.9E-1}} $\pm$ 1.1E-1 \\ MSE: \textcolor{red}{\textbf{3.5E-1}} $\pm$ 1.2E-1 / \textcolor{blue}{\textbf{3.5E-1}} $\pm$ 1.3E-1} \\
\bottomrule
\end{tabular}
}
\end{table}
\end{landscape}

\begin{landscape}
\begin{table}[t]
\centering
\setlength{\tabcolsep}{3pt}
\caption{Aggregate mitigation error across matched bounded/unbounded model pairs for the W-state hardware benchmarks. Each cell reports bounded / unbounded values. Top line: matched-pair count $k$, with bracketed coverage percentages (bounded finite / unbounded finite / matched pairs), each normalized by the estimated maximum number of observations (15). Rows may remain unmatched due to optimization failure or finite-value filtering. Bold red values indicate the smallest bounded MAE/MSE within a benchmark column; bold blue values indicate the smallest unbounded MAE/MSE within a benchmark column. Second line: MAE $\pm$ standard deviation. Third line: MSE $\pm$ standard deviation.}
\label{tab:paired-bounded-unbounded-real-w-state}
\resizebox{\linewidth}{!}{%
\begin{tabular}{@{}p{2.9cm}cccccc@{}}
\toprule
 & W-state, $X^{\otimes 30}$ & W-state, $X^{\otimes 40}$ & W-state, $X^{\otimes 50}$ & W-state, $Z^{\otimes 30}$ & W-state, $Z^{\otimes 40}$ & W-state, $Z^{\otimes 50}$  \\
\cmidrule(lr){2-7}
Model family (hyperparameters) & $\lambda = \{1,1.3,1.6\}$ & $\lambda = \{1,1.3,1.6\}$ & $\lambda = \{1,1.3,1.6\}$ & $\lambda = \{1,1.3,1.6\}$ & $\lambda = \{1,1.3,1.6\}$ & $\lambda = \{1,1.3,1.6\}$  \\
\midrule

Polynomial ($d=1$) & \makecell[l]{$k$=15 (100.00\% / 100.00\% / 100.00\%) \\ MAE: 3.1E-2 $\pm$ 2.1E-2 / \textcolor{blue}{\textbf{3.1E-2}} $\pm$ 2.1E-2 \\ MSE: 1.4E-3 $\pm$ 1.6E-3 / \textcolor{blue}{\textbf{1.4E-3}} $\pm$ 1.6E-3} & \makecell[l]{$k$=15 (100.00\% / 100.00\% / 100.00\%) \\ MAE: \textcolor{red}{\textbf{2.4E-2}} $\pm$ 1.5E-2 / \textcolor{blue}{\textbf{2.4E-2}} $\pm$ 1.5E-2 \\ MSE: \textcolor{red}{\textbf{7.6E-4}} $\pm$ 8.7E-4 / \textcolor{blue}{\textbf{7.6E-4}} $\pm$ 8.7E-4} & \makecell[l]{$k$=15 (100.00\% / 100.00\% / 100.00\%) \\ MAE: \textcolor{red}{\textbf{2.2E-2}} $\pm$ 1.5E-2 / \textcolor{blue}{\textbf{2.2E-2}} $\pm$ 1.5E-2 \\ MSE: \textcolor{red}{\textbf{6.9E-4}} $\pm$ 7.9E-4 / \textcolor{blue}{\textbf{6.9E-4}} $\pm$ 7.9E-4} & \makecell[l]{$k$=15 (100.00\% / 100.00\% / 100.00\%) \\ MAE: 6.7E-1 $\pm$ 3.8E-2 / 6.7E-1 $\pm$ 3.8E-2 \\ MSE: 4.5E-1 $\pm$ 5.1E-2 / 4.5E-1 $\pm$ 5.1E-2} & \makecell[l]{$k$=15 (100.00\% / 100.00\% / 100.00\%) \\ MAE: 8.8E-1 $\pm$ 2.3E-2 / 8.8E-1 $\pm$ 2.3E-2 \\ MSE: 7.8E-1 $\pm$ 4.1E-2 / 7.8E-1 $\pm$ 4.1E-2} & \makecell[l]{$k$=15 (100.00\% / 100.00\% / 100.00\%) \\ MAE: 9.8E-1 $\pm$ 2.6E-2 / 9.8E-1 $\pm$ 2.6E-2 \\ MSE: 9.5E-1 $\pm$ 5.1E-2 / 9.5E-1 $\pm$ 5.1E-2} \\
\midrule
Polynomial ($d=2$) & \makecell[l]{$k$=15 (100.00\% / 100.00\% / 100.00\%) \\ MAE: 1.9E-1 $\pm$ 1.8E-1 / 1.9E-1 $\pm$ 1.8E-1 \\ MSE: 6.7E-2 $\pm$ 1.1E-1 / 6.7E-2 $\pm$ 1.1E-1} & \makecell[l]{$k$=15 (100.00\% / 100.00\% / 100.00\%) \\ MAE: 1.4E-1 $\pm$ 9.0E-2 / 1.4E-1 $\pm$ 9.0E-2 \\ MSE: 2.8E-2 $\pm$ 3.1E-2 / 2.8E-2 $\pm$ 3.1E-2} & \makecell[l]{$k$=15 (100.00\% / 100.00\% / 100.00\%) \\ MAE: 1.4E-1 $\pm$ 9.8E-2 / 1.4E-1 $\pm$ 9.8E-2 \\ MSE: 2.8E-2 $\pm$ 3.3E-2 / 2.8E-2 $\pm$ 3.3E-2} & \makecell[l]{$k$=15 (100.00\% / 100.00\% / 100.00\%) \\ MAE: 3.6E-1 $\pm$ 2.5E-1 / 3.6E-1 $\pm$ 2.5E-1 \\ MSE: 1.9E-1 $\pm$ 2.1E-1 / 1.9E-1 $\pm$ 2.1E-1} & \makecell[l]{$k$=15 (100.00\% / 100.00\% / 100.00\%) \\ MAE: 6.6E-1 $\pm$ 2.4E-1 / 6.6E-1 $\pm$ 2.4E-1 \\ MSE: 4.9E-1 $\pm$ 2.9E-1 / 4.9E-1 $\pm$ 2.9E-1} & \makecell[l]{$k$=15 (100.00\% / 100.00\% / 100.00\%) \\ MAE: 9.5E-1 $\pm$ 1.8E-1 / 9.5E-1 $\pm$ 1.8E-1 \\ MSE: 9.4E-1 $\pm$ 3.2E-1 / 9.4E-1 $\pm$ 3.2E-1} \\
\midrule
Exponential ($a=0$) & \makecell[l]{$k$=15 (100.00\% / 100.00\% / 100.00\%) \\ MAE: 1.6E-1 $\pm$ 2.9E-1 / 5.0E+4 $\pm$ 1.3E+5 \\ MSE: 1.0E-1 $\pm$ 2.7E-1 / 1.8E+10 $\pm$ 5.1E+10} & \makecell[l]{$k$=15 (100.00\% / 100.00\% / 100.00\%) \\ MAE: 1.9E-1 $\pm$ 3.5E-1 / 2.0E+5 $\pm$ 5.1E+5 \\ MSE: 1.5E-1 $\pm$ 3.5E-1 / 2.8E+11 $\pm$ 7.3E+11} & \makecell[l]{$k$=15 (100.00\% / 100.00\% / 100.00\%) \\ MAE: 1.5E-1 $\pm$ 2.8E-1 / 2.8E+4 $\pm$ 1.0E+5 \\ MSE: 9.6E-2 $\pm$ 2.6E-1 / 1.1E+10 $\pm$ 4.2E+10} & \makecell[l]{$k$=4 (26.67\% / 100.00\% / 26.67\%) \\ MAE: 3.0E-1 $\pm$ 2.2E-1 / \textcolor{blue}{\textbf{3.0E-1}} $\pm$ 1.6E-1 \\ MSE: \textcolor{red}{\textbf{1.2E-1}} $\pm$ 1.2E-1 / \textcolor{blue}{\textbf{1.1E-1}} $\pm$ 1.1E-1} & \makecell[l]{$k$=8 (53.33\% / 100.00\% / 53.33\%) \\ MAE: 4.7E-1 $\pm$ 1.5E-1 / 5.8E-1 $\pm$ 2.7E-1 \\ MSE: 2.4E-1 $\pm$ 1.4E-1 / 4.0E-1 $\pm$ 2.8E-1} & \makecell[l]{$k$=11 (73.33\% / 100.00\% / 73.33\%) \\ MAE: 9.5E-1 $\pm$ 1.1E-1 / 1.4E+5 $\pm$ 4.7E+5 \\ MSE: 9.2E-1 $\pm$ 1.8E-1 / 2.2E+11 $\pm$ 7.3E+11} \\
\midrule
Exponential ($a$ is unconstrained) & \makecell[l]{$k$=3 (100.00\% / 20.00\% / 20.00\%) \\ MAE: \textcolor{red}{\textbf{2.5E-2}} $\pm$ 1.3E-2 / 8.9E+0 $\pm$ 1.3E+1 \\ MSE: \textcolor{red}{\textbf{7.4E-4}} $\pm$ 5.8E-4 / 2.0E+2 $\pm$ 3.4E+2} & \makecell[l]{$k$=3 (100.00\% / 20.00\% / 20.00\%) \\ MAE: 4.1E-2 $\pm$ 2.0E-2 / 5.4E-1 $\pm$ 8.5E-1 \\ MSE: 2.0E-3 $\pm$ 1.7E-3 / 7.7E-1 $\pm$ 1.3E+0} & \makecell[l]{$k$=2 (100.00\% / 13.33\% / 13.33\%) \\ MAE: 3.4E-2 $\pm$ 1.3E-2 / 6.6E-1 $\pm$ 2.2E-2 \\ MSE: 1.3E-3 $\pm$ 9.0E-4 / 4.4E-1 $\pm$ 2.9E-2} & \makecell[l]{$k$=6 (53.33\% / 86.67\% / 40.00\%) \\ MAE: 3.9E-1 $\pm$ 2.4E-1 / 9.1E-1 $\pm$ 1.2E+0 \\ MSE: 2.0E-1 $\pm$ 1.6E-1 / 2.0E+0 $\pm$ 4.4E+0} & \makecell[l]{$k$=5 (53.33\% / 60.00\% / 33.33\%) \\ MAE: 7.3E-1 $\pm$ 3.6E-1 / 7.0E-1 $\pm$ 1.6E-1 \\ MSE: 6.3E-1 $\pm$ 3.5E-1 / 5.1E-1 $\pm$ 2.2E-1} & \makecell[l]{$k$=1 (93.33\% / 6.67\% / 6.67\%) \\ MAE: 9.5E-1 $\pm$ 0.0E+0 / \textcolor{blue}{\textbf{7.1E-1}} $\pm$ 0.0E+0 \\ MSE: 9.1E-1 $\pm$ 0.0E+0 / \textcolor{blue}{\textbf{5.0E-1}} $\pm$ 0.0E+0} \\
\midrule
Poly-exp ($a=0$, $d=1$) & \makecell[l]{$k$=14 (93.33\% / 100.00\% / 93.33\%) \\ MAE: 1.8E-1 $\pm$ 3.4E-1 / 5.3E+4 $\pm$ 1.3E+5 \\ MSE: 1.4E-1 $\pm$ 3.2E-1 / 1.9E+10 $\pm$ 5.3E+10} & \makecell[l]{$k$=13 (86.67\% / 100.00\% / 86.67\%) \\ MAE: 1.5E-1 $\pm$ 3.0E-1 / 1.2E+5 $\pm$ 4.1E+5 \\ MSE: 1.1E-1 $\pm$ 2.8E-1 / 1.7E+11 $\pm$ 6.1E+11} & \makecell[l]{$k$=15 (100.00\% / 100.00\% / 100.00\%) \\ MAE: 1.4E-1 $\pm$ 2.8E-1 / 2.8E+4 $\pm$ 1.0E+5 \\ MSE: 9.5E-2 $\pm$ 2.6E-1 / 1.1E+10 $\pm$ 4.2E+10} & \makecell[l]{$k$=4 (26.67\% / 100.00\% / 26.67\%) \\ MAE: \textcolor{red}{\textbf{3.0E-1}} $\pm$ 2.2E-1 / \textcolor{blue}{\textbf{3.0E-1}} $\pm$ 1.6E-1 \\ MSE: 1.2E-1 $\pm$ 1.2E-1 / \textcolor{blue}{\textbf{1.1E-1}} $\pm$ 1.1E-1} & \makecell[l]{$k$=9 (60.00\% / 100.00\% / 60.00\%) \\ MAE: \textcolor{red}{\textbf{4.3E-1}} $\pm$ 1.9E-1 / \textcolor{blue}{\textbf{5.3E-1}} $\pm$ 3.1E-1 \\ MSE: \textcolor{red}{\textbf{2.2E-1}} $\pm$ 1.6E-1 / \textcolor{blue}{\textbf{3.6E-1}} $\pm$ 2.9E-1} & \makecell[l]{$k$=13 (86.67\% / 100.00\% / 86.67\%) \\ MAE: 8.1E-1 $\pm$ 3.5E-1 / 1.2E+5 $\pm$ 4.3E+5 \\ MSE: 7.7E-1 $\pm$ 3.9E-1 / 1.9E+11 $\pm$ 6.8E+11} \\
\midrule
Poly-exp ($a=0$, $d=2$) & \makecell[l]{$k$=14 (93.33\% / 100.00\% / 93.33\%) \\ MAE: 5.8E-2 $\pm$ 1.3E-1 / 1.7E+67 $\pm$ 6.4E+67 \\ MSE: 1.9E-2 $\pm$ 6.4E-2 / 4.1E+135 $\pm$ 1.5E+136} & \makecell[l]{$k$=14 (93.33\% / 100.00\% / 93.33\%) \\ MAE: 4.1E-2 $\pm$ 6.6E-2 / 1.1E+69 $\pm$ 4.1E+69 \\ MSE: 5.8E-3 $\pm$ 1.3E-2 / 1.6E+139 $\pm$ 6.1E+139} & \makecell[l]{$k$=15 (100.00\% / 100.00\% / 100.00\%) \\ MAE: 6.4E-2 $\pm$ 1.9E-1 / 2.6E+65 $\pm$ 1.0E+66 \\ MSE: 3.7E-2 $\pm$ 1.4E-1 / 1.0E+132 $\pm$ 3.9E+132} & \makecell[l]{$k$=9 (60.00\% / 100.00\% / 60.00\%) \\ MAE: 9.0E-1 $\pm$ 1.2E-1 / 8.9E-1 $\pm$ 1.5E-1 \\ MSE: 8.2E-1 $\pm$ 2.0E-1 / 8.1E-1 $\pm$ 2.3E-1} & \makecell[l]{$k$=11 (73.33\% / 100.00\% / 73.33\%) \\ MAE: 7.7E-1 $\pm$ 1.5E-1 / 1.7E+9 $\pm$ 5.5E+9 \\ MSE: 6.1E-1 $\pm$ 2.2E-1 / 3.0E+19 $\pm$ 1.0E+20} & \makecell[l]{$k$=15 (100.00\% / 100.00\% / 100.00\%) \\ MAE: 9.0E-1 $\pm$ 2.2E-1 / 6.7E+68 $\pm$ 2.6E+69 \\ MSE: 8.6E-1 $\pm$ 2.7E-1 / 6.6E+138 $\pm$ 2.5E+139} \\
\midrule
Poly-exp ($a$ is unconstrained, $d=1$) & \makecell[l]{$k$=3 (100.00\% / 20.00\% / 20.00\%) \\ MAE: 4.6E-1 $\pm$ 3.7E-1 / 8.9E+0 $\pm$ 1.3E+1 \\ MSE: 3.0E-1 $\pm$ 2.8E-1 / 2.0E+2 $\pm$ 3.4E+2} & \makecell[l]{$k$=3 (100.00\% / 20.00\% / 20.00\%) \\ MAE: 3.7E-1 $\pm$ 5.5E-1 / 5.4E-1 $\pm$ 8.5E-1 \\ MSE: 3.4E-1 $\pm$ 5.8E-1 / 7.7E-1 $\pm$ 1.3E+0} & \makecell[l]{$k$=2 (100.00\% / 13.33\% / 13.33\%) \\ MAE: 4.5E-1 $\pm$ 1.4E-1 / 6.6E-1 $\pm$ 2.2E-2 \\ MSE: 2.1E-1 $\pm$ 1.3E-1 / 4.4E-1 $\pm$ 2.9E-2} & \makecell[l]{$k$=6 (53.33\% / 86.67\% / 40.00\%) \\ MAE: 3.6E-1 $\pm$ 2.2E-1 / 6.3E-1 $\pm$ 5.0E-1 \\ MSE: 1.7E-1 $\pm$ 1.4E-1 / 6.1E-1 $\pm$ 9.9E-1} & \makecell[l]{$k$=5 (60.00\% / 60.00\% / 33.33\%) \\ MAE: 5.1E-1 $\pm$ 4.1E-1 / 3.9E+0 $\pm$ 7.1E+0 \\ MSE: 4.0E-1 $\pm$ 3.6E-1 / 5.5E+1 $\pm$ 1.2E+2} & \makecell[l]{$k$=1 (93.33\% / 6.67\% / 6.67\%) \\ MAE: \textcolor{red}{\textbf{7.1E-1}} $\pm$ 0.0E+0 / \textcolor{blue}{\textbf{7.1E-1}} $\pm$ 0.0E+0 \\ MSE: \textcolor{red}{\textbf{5.1E-1}} $\pm$ 0.0E+0 / \textcolor{blue}{\textbf{5.0E-1}} $\pm$ 0.0E+0} \\
 \bottomrule
\end{tabular}
}
\end{table}
\end{landscape}

\end{document}